\documentclass[preprint,12pt,3p]{elsarticle}

\usepackage{graphicx}
\usepackage{amssymb}
\usepackage{caption}
\usepackage{url}
\usepackage{amsmath}
\usepackage{float}
\usepackage{xcolor}
\usepackage{lineno}
\usepackage{ulem}
\usepackage{mathrsfs}  
\usepackage[mathscr]{euscript}
\usepackage{euscript}
\usepackage[mathscr]{euscript}
\usepackage{multirow}

\usepackage{tikz}
\usetikzlibrary{shapes.geometric, arrows}
\usetikzlibrary{positioning}
\tikzstyle{arrow} = [thick,->,>=stealth]

\def \f{\vec{f}}

\def\f {{{\bf f}}}
\def\u {{{\bf u}}}

\def\div {{\nabla \cdot}}

\newcommand{\beginsupplement}{%
        \setcounter{table}{0}
        \renewcommand{\thetable}{S\arabic{table}}%
        \setcounter{figure}{0}
        \renewcommand{\thefigure}{S\arabic{figure}}%
        \setcounter{equation}{0}
        \renewcommand{\theequation}{S\arabic{equation}}%
        \setcounter{section}{0}
        \renewcommand{\thesection}{S\arabic{section}}%
     }

\graphicspath{{./BB_Images/}}
\graphicspath{{./Images/}}

\journal{IOPscience}

\begin{document}

\begin{frontmatter}


\title{Hopscotching Jellyfish: combining different duty cycle kinematics can lead to enhanced swimming performance}

 \author[label1]{Tierney Baldwin}
 \address[label1]{Dept. of Mathematics and Statistic, 2000 Pennington Road, The College of New Jersey, Ewing Township, NJ 08628, USA}

\author[label1]{Nicholas A. Battista}
\cortext[cor1]{I am corresponding author}
\ead{battistn@tcnj.edu}


\begin{abstract}

Jellyfish (Medusozoa) have been deemed the most energy-efficient animals in the world. Their bell morphology and relatively simple nervous systems make them attractive to robotocists. Although, the science community has devoted much attention to understanding their swimming performance, there is still much to be learned about the jet propulsive locomotive gait displayed by prolate jellyfish. Traditionally, computational scientists have assumed uniform duty cycle kinematics when computationally modeling jellyfish locomotion. In this study we used fluid-structure interaction modeling to determine possible enhancements in performance from shuffling different duty cycles together across multiple Reynolds numbers and contraction frequencies. Increases in speed and reductions in cost of transport were observed as high as 80\% and 50\%, respectively. Generally, the net effects were greater for cases involving lower contraction frequencies. Overall, robust duty cycle combinations were determined that led to enhanced or impeded performance. 

\end{abstract}


\date{\today}


\end{frontmatter}

%
%

%
%

%
%

%
%

\section{Introduction}
\label{sec:intro}

Across the animal kingdom there are a variety of modes of locomotion and swimming gaits in aquatic organisms. Different modes may be more effective (and efficient) at particular fluid scales (i.e., Reynolds numbers, Re) than others \cite{Vogel:1996}. At intermediate Reynolds numbers ($\mathrm{Re}\sim O(1)-O(100)$) and higher  ($\mathrm{Re}\gtrsim O(1000)$), some undulatory locomotion modes include the anguilliform swimming of lampreys and eels \cite{Tytell:2004}, the carangiform swimming in most fishes \cite{Lucas:2020}, and thunniform swimming in tuna \cite{Shadwick:2008} and marine mammals like whales and dolphins \cite{Williams:2009}. However, many other organisms use propulsatory modes, either drag-based mechanisms, like that of krill \cite{Murphy:2013}, suction based mechanisms like those of oblate jellyfish \cite{Gemmell:2015}, or momentum injection like prolate jellyfish \cite{Costello:2008,Costello:2020}. The scientific community has given great attention to studying the performance of such modes, either through organismal experiments, \cite{Bainbridge:1958,Dabiri:2006,Dabiri:2010b,Flammang:2011,Flammang:2013}, physical experiments \cite{Lauder:2012,Floryan:2017,Ford:2020}, reduced-order modeling \cite{Daniel:1983,McHenry:2003}, or computational fluid dynamics (CFD) models \cite{Park:2014,Akoz:2017,Wang:2018,Hamlet:2018,Hoover:2018,Battista:ICB2020,Hoover:2020,Miles:2019}. Beyond basic science, one motivation for all of these efforts is to inform the design of novel biomimetic autonomous underwater vehicles \cite{Roper:2011,Costa:2018}.

To that extent, bioinspired computational studies are invaluable in this process as they allow scientists to more easily probe either biological or theoretical parameter spaces, in a much reduced cost-efficient and time-efficient manner. For example, widespread morphological and/or varying kinematic studies can find parameter combinations that lead to optimal swimming for a given performance metric \cite{Kern:2006,VanRees:2015,Tokic:2019}. Ultimately, performance (speed and cost of transport) are observed to be highly sensitive to kinematic variations. For a given morphology, performance was found to be the most sensitive to frequency in a general anguilliform and jet propulsion study \cite{Battista:ICB2020b,Miles:2019}, even illustrating nonlinear dependence on frequency. On the other hand, fish-like locomotive modes have demonstrated linear relationships between speed and frequency \cite{Bainbridge:1958,Hunter:1971,Videler:1991}. However, kinematic variations that go beyond changing a swimmer's frequency have been shown to be important. 

For example, the transition from continuous to intermittent swimming has been investigated for fish-like locomotion modes. In continuous swimming a fish's caudal fin continuously beats, while in intermittent swimming, there are pauses at the end of each symmetric tail beat. Intermittent swimming, also known as \textit{burst and glide} (or burst and coast) swimming, was theoretically predicted to substantially decrease the energetic cost of swimming a particular distance by upwards of 50\% \cite{Lighthill:1971,Weihs:1974}. CFD models have confirmed such predictions \cite{Akoz:2017,Wang:2018,Zhao:2020} and identified transitions in which it becomes energetically favorable to move from a continuous or intermittent mode \cite{Han:2020}. 

Similar burst and glide behavior has been observed in prolate jellyfish, i.e., jellyfish whose bell height to width ratio is greater than 1. These jellyfish generally have longer pauses after bell relaxation (expansion) and thereby take advantage of greater contributions in forward swimming from \textit{passive energy recapture} \cite{Gemmell:2018}. Passive energy recapture is a mechanism which occurs when a jellyfish pauses any active bell motion after its expansion phase, resulting in additional growth of the stopping vortex inside their bell. The strength of this stopping vortex determines how much additional flow is directed towards the umbrellar cavity of the bell, creating more momentum that favors the direction of swimming \cite{Gemmell:2013}. Passive energy recapture is common among jellyfish; however, contributions by it vary among taxa. Although, the magnitude of contribution by PER does appear to be determined by morphology or swimming speeds, but instead more heavily related to pause duration \cite{Gemmell:2018}.   

A useful parameter to study burst and glide behavior is the \textit{duty cycle}. The duty cycle can be quantified as the ratio of total burst period to total cycle period. Previous mathematical models of jet propulsion in jellyfish have used similar definitions \cite{McHenry:2003,Miles:2019}, where the \textit{burst} is considered the period in which the bell is actively contracting. Thus, the remaining time of the cycle involved either bell expansion or pauses between successive pulses. Both of these studies viewed the duty cycle as a prescribed parameter in the model, while other computational studies model contractions via an emergent process of applying tension to the bell itself \cite{Hoover:2017,Hoover:2019,Hoover:2021}.

However, across all of these simulations, the duty cycle and overall contraction kinematics did not vary from cycle to cycle. In CFD studies of aquatic locomotion this is a common assumption and aspect in each study.  We posit that combining different duty cycles could lead to enhanced swimming performance for jet propulsion modes of locomotion. These enhancements are believed to arise from differences in the underlying cycle-to-cycle fluid dynamics created by changing duty cycles, i.e., vortex formation, persistence, and interactions from the previous cycle to the next.

To test this hypothesis, we used the two-dimensional jellyfish fluid-structure interaction model of Hoover and Miller 2015 \cite{Hoover:2015} and Miles and Battista \cite{Miles:2019} to investigate how stitching different duty cycles combinations together may boost (or impede) swimming performance. In particular, we wanted to address how optimal combinations may vary across multiple fluid scales ($\mathrm{Re}=37.5,75,150$, and $300$) and bell actuation (stroke) frequencies ($f=0.5, 0.75, 1.125, 1.5$, and $1.875\ \mathrm{Hz}$). As stroke frequency was previously shown to be the most important parameter affecting performance at these fluid scales \cite{Miles:2019}, we expected that optimal duty cycle combinations would vary more across frequency than Reynolds numbers.

%
%

%
%
%
%
%
%

\section{Mathematical Methods}
\label{sec:methods}

An immersed boundary method (IB) was used to solve the fluid-structure interaction problem involving a flexible jellyfish bell immersed in a viscous, incompressible fluid. The IB software, \texttt{IB2d} \cite{Battista:2015,BattistaIB2d:2017,BattistaIB2d:2018}, was used, whose IB implementation is based off of the original by Charles Peskin \cite{Peskin:1972,Peskin:1977,Peskin:2002}. See the Supplemental Materials for more details on IB. The jellyfish model explored here is one of the models available upon download (or clone) of \texttt{IB2d} \cite{Miles:2019}:

\begin{center}
\texttt{IB2d$/$matIB2d$/$Examples$/$Examples$\_$Jellyfish$\_$Swimming$/$Hoover$\_$Jellyfish}.
\end{center}

The IB is still a leading numerical framework for studying problems in FSI due to its robustness, simplicity, and flexibility in modelling complex deformable structures, like those that arise in many biological contexts \cite{BattistaIB2d:2017,BattistaIB2d:2018}. Moreover, it has been improved upon numerous times since its initial conception \cite{Fauci:1993,Lai:2000,Griffith:2005,Mittal:2005,Griffith:2007,BGriffithIBAMR,Griffith:IBFE}. It has been previously used in numerous applications ranging from physiological flows, such as those in cardiac fluid dynamics \cite{Griffith:2012b,Battista:2017,Lee:2020}, blood vessels \cite{Kim:2009}, lymphatic capillaries \cite{Senter:2020}, to aquatic locomotion \cite{Fauci:1988,Bhalla:2013a,Hamlet:2015,Zhang:2018} to animal flight \cite{Miller:2005,Windes:2018} to flow past biological structures, such as vegetation \cite{Chen:2019}, leaves \cite{Miller:2012}, seeds \cite{Kim:2016}, corals \cite{Hossain:2020}, or biofilms \cite{Nguyen:2021_IB2d}.

In the remainder of this section we will briefly introduce the computational geometry and kinematic model governing the contraction dynamics. The computational model is based on that of Hoover and Miller 2015 \cite{Hoover:2015}, which was extensively further explored using dynamic similarity by Miles and Battista 2020 \cite{Miles:2019}. Miles and Battista 2020 \cite{Miles:2019} used it to assess its swimming performance's global sensitivity across an input parameter space consisting of Re, frequency, and duty cycle. Thus, the computational geometry and numerical parameters are identical to \cite{Miles:2019}, although the contraction kinematics are altered to include combinations of two duty cycles. The governing equations of IB are provided in the Supplemental Materials.

%
%

\subsection{Computational Geometry}
\label{sec:jelly_model}

\begin{figure}[H]
\centering
\includegraphics[width=0.65\textwidth]{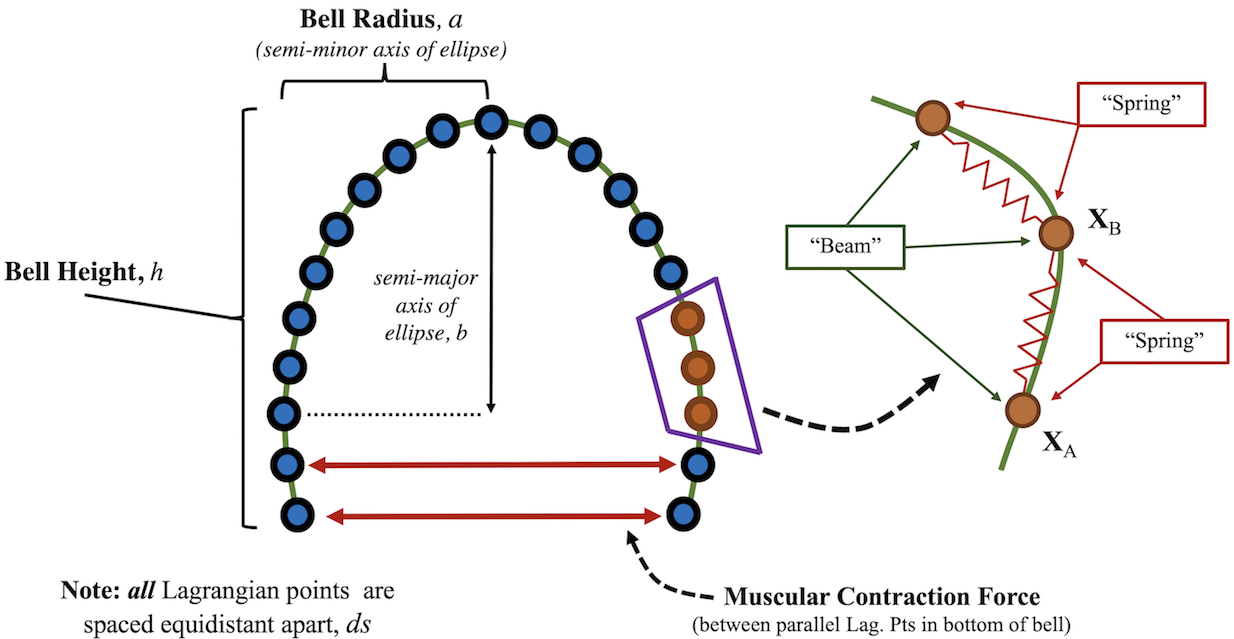}
\caption{The Hoover and Miller 2015 \cite{Hoover:2015} jellyfish model geometry uses a semi-elliptical geometry. The points are connected by virtual springs and virtual non-invariant beams in both the original model implemented in \textit{IBAMR} \cite{BGriffithIBAMR} and this study implemented in \textit{IB2d}. Image courtesy of \cite{Miles:2019}.}
\label{fig:Model_Geometry}
\end{figure}

The jellyfish's morphology was identical to that of Hoover and Miller 2015 \cite{Hoover:2015}, i.e., the bell was modeled as a semi-ellipse with semi-major axis, $b=0.75$, and semi-minor axis, $a=0.5$. The bell encompassed more than the top hemi-ellipse, see Figure \ref{fig:Model_Geometry}. Its overall height was $h=1.0$. As the jellyfish bell's geometry was conserved across every simulation performed, it was chosen such that the bell's fineness ratio was exactly equal to $1.0$. The fineness ratio, $FR$, is the ratio of the bell height its diameter \cite{Dabiri:2007}. Therefore we explored how different duty cycles may affect performance directly at the interface between oblate and prolate jellyfish. These jellyfish have fineness ratios less than one and greater than one, respectively. However, our model's jet propulsive gait more closely depicts prolate jellyfish, as they use jet propulsion. Oblate jellyfish  use a different locomotion mode, a suction-based jet-paddling propulsion mode \cite{Dabiri:2007,Gemmell:2015}.

The bell was discretized into equidistant points, a distance of $ds$ apart, and was modeled in a Lagrangian frame of reference. The \textit{Lagrangian} grid resolution was twice as resolved as the background fluid grid, i.e., $ds=0.5dx$, which is described in an \textit{Eulerian} frame. This resolution choice is traditionally made in IB to minimize interpolation errors between the Eulerian and Lagrangian grids \cite{Peskin:2002}. 

All communication between the Eulerian (fluid) and Lagrangian (jellyfish) meshes are handled through integral equations with delta function kernels (see Eqs. \ref{eq:force1}-\ref{eq:force2} in the Supplemental Materials for more details). Such integrals provide a mechanism to interpolate data between grids. That is, when forces are spread from the jellyfish onto the fluid, it is done so at the points nearest the jellyfish itself. Moreover, through the numerical delta functions used, the fluid grid points nearest the jellyfish feel the brunt of such force, while grid points further away feel less. Similarly, when velocity information is interpolated back onto the jellyfish (in order to move it at the local fluid velocity), it is only the velocity data from the fluid grid points near the jellyfish to which are used, with the closest affecting the motion to a greater extent. This framework makes it possible to allow the jellyfish to move around and deform, without restricting it to specific grid locations, i.e., it can be modeled using a time-dependent curvilinear Lagrangian mesh. Therefore, the fluid equations can be elegantly solved on a fixed Cartesian mesh (see Eqs. \ref{eq:NS1}-\ref{eq:NSDiv1} in the Supplemental Materials).

Each point along the semi-elliptical bell were connected by \textit{virtual springs} and \textit{virtual (non-invariant) beams} in the \textit{IB2d} framework, as shown in Figure \ref{fig:Model_Geometry}. A virtual spring between two Lagrangian points governs how much the two points may stretch apart or compress towards each other. On the other hand, a virtual beam governs the bending allowance between three successive points. Both the springs and beams have a preferred resting length or curvature, respectively. If there is a perturbation that causes the configuration to deviate away from its preferred state, there is a restorative deformation force to drive the geometry back. Therefore, its preferred configuration is also the model's lowest energy state. In our jellyfish model, this is when the bell is initially at rest, when it is neither expanding nor contracting. One way in which to compute such deformation forces is below:
\begin{align}
\label{eq:spring} &\text{\footnotesize $\textbf{F}_{spr} = k_{spr} \left( 1 - \frac{R_L(t)}{||{\bf{X}}_{A}(t)-{\bf{X}}_{B}(t)||} \right) \cdot \left( \begin{array}{c} x_{A}(t) - x_{B}(t) \\ y_{A}(t) - y_{B}(t) \end{array} \right) $}\\
    \label{eq:beam} &\textbf{F}_{beam} = k_{beam} \frac{\partial^4}{\partial s^4} \Big( \textbf{X}_C(t) - \textbf{X}^{con} \Big),
\end{align}
where (\ref{eq:spring}) and (\ref{eq:beam}) model deformation forces for springs and beams, respectively. The spring and beam stiffnesses, $k_{spr}$ and $k_{beam}$, respectively, allow us to appropriately penalize deviations from the jellyfish's preferred geometric configuration. In (\ref{eq:spring}), $R_L(t)$ denotes a possible time-dependent resting length and the quantities $\textbf{X}_{A}=\langle x_A,y_A\rangle$ and $\textbf{X}_{B}=\langle x_B,y_B\rangle$ denote two Lagrangian points tethered together by a virtual spring (see Figure \ref{fig:Model_Geometry}). Note that along the jellyfish bell that $R_L(t)=ds$ for all spring connections and that $k_{spr}$ was large to minimize both stretching and compression of the bell. In (\ref{eq:beam}),  $\textbf{X}_C(t)$ and $\textbf{X}^{con}$ denote a Lagrangian point on the interior of the bell and its corresponding initial (preferred) geometric curvature, respectively. Modeling the jellyfish bell with beams allowed the bell to bend and therefore contract for jet propulsive purposes. The $4^{th}$-order derivative discretization for the beams can be found in  \cite{BattistaIB2d:2018}.

%
%

\subsection{Contraction Kinematics}
\label{sec:methods_kinematics}

Virtual springs were used to mimic the subumbrellar and coronal muscles that induce bell contractions, via dynamically changing their resting lengths. These springs tethered points across the jellyfish bell that were below the top hemi-ellipse. While the deformation force equation is identical to Eq. (\ref{eq:spring}), a different spring stiffness coefficient, called $k_{muscle}$, and time-dependent spring resting lengths, $R_L(\tilde{t})$, were used. Each simulation defined the period of one complete bell actuation cycle, $T$, based off its specified actuation (stroke) frequency, $f$. Furthermore, we were able to integrate different duty cycle combinations through an appropriate choice of the functional form of the time-dependent resting lengths.

First, we used modular arithmetic to define $\tilde{t}= t \mod 2T$, so that $0\leq\tilde{t}\leq 2T$, based on a particular actuation frequency, i.e., $T=1/f$. Next we chose a specific duty cycle combination $(p_1,p_2)$, where $p_1$ and $p_2$ were the duty cycles for the first and second contraction cycles, respectively. Hence rather than only repeat one duty cycle over and over, this two-combination of duty cycles would be repeated throughout the simulation. Recall that we defined the duty cycle to be the overall percentage of the actuation cycle in which the bell is actively contracting. Therefore, the overall fraction in which the bell was expanding would be $(1-p_1)T$ and $(1-p_2)T$, respectively. For this preliminary study involving duty cycle two-combinations, the bell would only ever be actively contracting or expanding, with no rest between successive actuation cycles. 

\begin{figure}[H]
\centering
\includegraphics[width=0.7\textwidth]{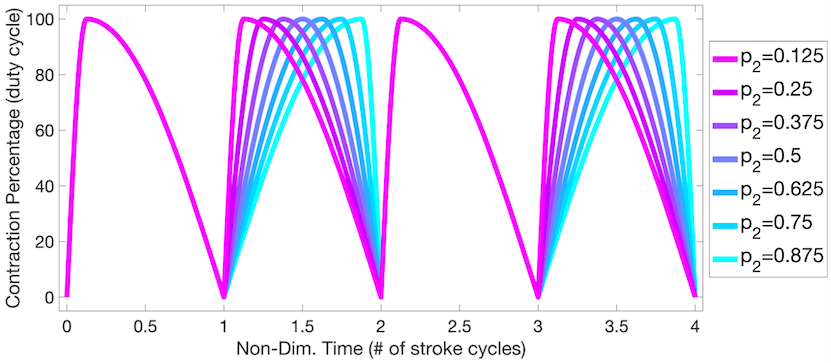}
\caption{An illustration of different duty cycle combinations $(p_1,p_2)$ with uniform $p_1=0.125$.}
\label{fig:Duty_Cycle_Exmp}
\end{figure}

Hence we could mathematically represent $R_L(\tilde{t})$  in the following manner:

\begin{equation}
    \label{eq:RL_muscle} R_L(\tilde{t}) = \left\{\begin{array}{cc} 
        2a\cos\left( \frac{\pi \tilde{t} }{2p_1T} \right) & \ \  \tilde{t}\leq p_1T  \\
        2a\sin\left( \frac{\pi}{2(1-p_1)T}\left(\tilde{t}-p_1T\right) \right) &\ \ p_1T< \tilde{t}\leq T  \\
        2a\cos\left( \frac{\pi (\tilde{t}-T) }{2p_2T} \right) & \ \  T < \tilde{t}\leq (1+p_2)T   \\
        2a\sin\left( \frac{\pi}{2(1-p_2)T}\left(\tilde{t}-(1+p_2)T\right) \right) &\ \ (1+p_2)T < \tilde{t}\leq 2T 
    \end{array}\right.
\end{equation}
An example of such duty cycle is provided in Figure \ref{fig:Duty_Cycle_Exmp}. This figure illustrates possible duty cycle combinations where $p_1=0.125$ and $p_2$ is varied, for the first 4 actuation cycles. Since we are exploring combinations of two duty cycles, the contraction kinematics are identical in cycle 1 and 3, since all have the same $p_1$, while they are different in cycle 2 and 4, due to differing $p_2$. We note that prolate jellyfish generally have smaller duty cycles (they contract on faster time scales than their expansion phase) \cite{Ford:2000,Colin:2013,Kakani:2013}; however, there is variation. Approximate duty cycles for some species of jellyfish with fineness ratios of $\sim1$ are given in Table \ref{table:duty_cycles}. Notably, depending on the size scale of the jellyfish, it may maintain approximately the same fineness ratio, but have substantially different duty cycle \cite{Colin:2013}, see \textit{Chiropsella bronzie} in Table \ref{table:duty_cycles}, whose duty cycles were observed to be either $\sim0.13$ or $0.31$ whether it was `small' or `large', respectively. Therefore since scale ($\mathrm{Re}$) may play a role as to what duty cycle a jellyfish has, considering different combinations of $\mathrm{Re}$ and duty cycle is interesting to explore. Overall, generally lower duty cycles more realistically depict prolate jellyfish kinematics. However, in order to thoroughly explore this idea of duty cycle combinations, we considered more cases than what are biologically realistic, i.e., $$p_1,p_2\in\{0.05,0.10,0.15,\ldots,0.50,\ldots,0.95 \}.$$

\begin{table}
\begin{center}
\resizebox{\textwidth}{!}{%
\begin{tabular}{| c | c | c | c | c | c | c | c |}
    \hline
    Species               & $D$ (cm) & $FR$  & Duty Cycle & $f$ ($s^{-1}$) & Re & Speed ($1/$St)  & Estimated from \\ \hline
    \textit{Aglanta digitale} &   0.8          & $2$  & $0.07$     & 1.43  & 91.5 & 1.91  &\cite{Colin:1996,Ford:2000,Colin:2002} \\ \hline
    \textit{Chrionex fleckeri} (small) & 2   & $0.925$   & $0.22$   & 1.81 & 727 & 0.72 & \cite{Colin:2013}           \\ \hline
    \textit{Chrionex fleckeri} (large) & 13.6   & $1$      & $0.31$ & 1.81 & 3360 & 0.23 & \cite{Colin:2013}           \\ \hline
    \textit{Chiropsella bronzie} (small) & 0.66 & $0.8$    & $0.13$ & 1.65 & 726 & 1.36 &\cite{Colin:2013}           \\ \hline
    \textit{Chiropsella bronzie} (large) & 6.1  & $0.8$    & $0.31$ & 1.33 & 4961 & 0.88 & \cite{Colin:2013}          \\ \hline
    \textit{Sarsia tubulosa} (small)    & 0.11  & $1.25$   & $0.41$ & 5.26 & 6 & 0.23 & \cite{Kakani:2013}     \\ \hline
    \textit{Sarsia tubulosa} (large)    & 0.85 & $1.07$   & $0.19$ & 0.91 & 65 & 2.13 & \cite{Colin:2002}     \\ \hline
    \hline
    \end{tabular}
    }
    \caption{The estimated fineness ratios, contraction frequencies, and duty cycles for four species of jellyfish, three of which include estimates for smaller and larger medusae. Their bell diameter ($D$) as well as estimated Reynolds number and average swimming speed are also given. The average swimming speeds are dimensionless and can be interpreted as bell diameters per contraction cycle.}
    \label{table:duty_cycles}
    \end{center}
\end{table}

%
%

\subsection{The input parameter space and performance metrics}
\label{sec:parameters}

In order to compare data across different frequencies and fluid scales, we used \textit{dynamic similarity}, i.e., dynamic scaling, when defining all of our dimensionless quantities. This was to help ensure proper comparative analysis across our input model parameter spaces. First, we defined the fluid scale by a frequency-based Reynolds number, $\mathrm{Re}$. The characteristic length was set to the bell diameter at rest, $D_{jelly}$, and the characteristic frequency was set to the bell's actuation (stroke) frequency, $f_{jelly}$. Therefore our characteristic velocity scale was set to $V_{jelly}=f_{jelly}D_{jelly}$. This gave the following definition of Re,
\begin{equation}
    \label{eq:Re_study} Re = \frac{f_{jelly} D_{jelly}^2 }{\nu}.
\end{equation}
Across all studies the fluid's kinematic viscosity, $\nu$, was selectively chosen to give a specific $\mathrm{Re}$ value, for a particular contraction frequency, $f$. From our use of dynamic scaling, we could use the same jellyfish geometry and material properties, as well as uniform computational parameters across all simulations performed, e.g., the stiffness parameters, $k_{beam}$, $k_{spr}$, and $k_{muscle}$, the grid size and resolution, and the time-step, $dt$. Table \ref{table:num_param} provides these computational parameters.

\begin{table}
\begin{center}
\begin{tabular}{| c | c | c | c |}
    \hline
    Parameter               & Variable    & Units        & Value \\ \hline
    Domain Size            & $[L_x,L_y]$  & m               &  $[5,24]$             \\ \hline
    Spatial Grid Size      & $dx=dy$      & m               &  $L_x/320=L_y/1536$            \\ \hline
    Lagrangian Grid Size    & $ds$        & m               &  $dx/2$               \\ \hline
    Time Step Size          & $dt$        & s               &  $10^{-5}$   \\ \hline
    Total Simulation Time    & $\mathscr{T}$  & \textit{pulses} &  $8$               \\ \hline
    Fluid Kinematic Viscosity & $\nu$      & $\mathrm{cm}^2/ \mathrm{s}$       &  \textit{varied}   \\ \hline
    Bell Radius           & $a$           & m       &  $0.5$   \\ \hline
    Bell Diameter           & $D$ ($2a$)  & m       &  $1.0$ \\ \hline
    Bell Height           & $h$           & m       &  $1.0$   \\ \hline
    Contraction Frequency   & $f$     & $\mathrm{Hz}$   &  \textit{varied}     \\ \hline
    Actuation Period   & $T=1/f$          & s          &  \textit{varied}     \\ \hline
    Duty Cycle Combinations & $(p_1,p_2)$  & &  \textit{varied}     \\ \hline
    Spring Stiffness   & $k_{spr}$ & $\mathrm{kg}\cdot \mathrm{m}/\mathrm{s}^2$ &  $1\times10^{7}$  \\ \hline
    Beam Stiffness   & $k_{beam} $ & $\mathrm{kg}\cdot \mathrm{m}/\mathrm{s}^2$ &  $2.5\times10^{5}$  \\ \hline
    Muscle Spring Stiffness & $k_{muscle}$ & $\mathrm{kg}\cdot \mathrm{m}/\mathrm{s}^2$ &  $1\times10^{5}$\\
    \hline
    \end{tabular}
    \caption{Numerical parameters used in the two-dimensional simulations.}
    \label{table:num_param}
    \end{center}
\end{table}

Periodic boundary conditions were used on all edges of the computational domain for all simulations performed. The computational width was kept constant with $L_x=5.$ Thorough convergence studies had been previously performed on this jellyfish model \cite{Miles:2019,BattistaMizuhara:2019}. Low relative errors were observed in swimming speed for computational domain widths of $L_x\in[3,8]$ for $Re=\{37.5,75,150,300\}$ \cite{Miles:2019}. Narrower computational domains led to slightly decreased forward swimming speeds while qualitative differences vortex formation were minimal, amongst all cases.
Grid resolution studies performed in \cite{BattistaMizuhara:2019} demonstrated convergence in both the Lagrangian (jellyfish) and Eulerian (fluid) data for the computational parameters given in Table \ref{table:num_param}.

The following data was stored equally spaced time points across each bell contraction cycle of each simulation:
\begin{enumerate}
    \item Position of Lagrangian Points: $\textbf{X}(s,t)$
    \item Horizontal/Vertical forces on each Lagrangian Point: $\textbf{F}(s,t)$
    \item Fluid Velocity: $\textbf{u}(\textbf{x},t)=\big(u(\textbf{x},t),v(\textbf{x},t)\big)$ 
    \item Fluid Vorticity: $\omega(\textbf{x},t)$
    \item Fluid Pressure: $P(\textbf{x},t)$
    \item Forces spread onto the Fluid (Eulerian) grid from the Jellyfish (Lagrangian) mesh: $\textbf{f}(\textbf{x},t)$
\end{enumerate}
We used the open-source visualization software VisIt \cite{HPV:VisIt}, created and maintained by Lawrence Livermore National Laboratory, and MATLAB \cite{MATLAB:2015a} for post-processing of all the simulation data. In this work we examine two key performance metrics - a dimensionless swimming speed and a power-based cost of transport ($COT$). In order to compute these metrics, we temporally-averaged the simulation data across the 7th and 8th actuation cycles. The jellyfish had reached steady swimming speeds by those cycles (see Figures \ref{fig:TimeEvo_Re75_VaryFreq} and \ref{fig:TimeEvo_f0pt75_VaryRe}).
The dimensionless forward swimming speed was given by the inverse of the Strouhal number, $\mathrm{St}$, which itself is defined by 
\begin{equation}
    \label{eq:Str} \mathrm{St} = \frac{fD}{V_{dim}}
\end{equation}
where $V_{dim}$, $f$, and $D$ are the \textit{dimensional} swimming speed, the bell's actuation frequency, and the maximum diameter during an actuation cycle. Thus, by taking the inverse of the Strouhal number as our dimensionless swimming speed, it provided a normalized forward swimming speed based on driving frequency and bell diameter, i.e., $1/\mathrm{St}=V_{dim}/(fD)$. It can also be interpreted as a speed in terms of bell diameters/cycle. Previous studies of animal locomotion have highlighted that $0.2 < St < 0.4$ correspond to high locomotion efficiency \cite{Triantafyllou:1991,Taylor:2003,Floryan:2018}. However, rather than use $\mathrm{St}$ to assess swimming efficiency, we computed a dimensionless cost of transport ($COT$). The cost of transport measures the input power (or energy) spent by the swimmer per unit speed of the swimmer itself \cite{Schmidt:1972,Bale:2014,Hamlet:2015}. We calculated $COT$ by the following equation

\begin{equation}
    \label{eq:COT_dim} COT = \frac{1}{N} \frac{1}{\rho f^2 D^3} \frac{1}{V_{dim}} \displaystyle\sum_{j=1}^N |F_j||U_j|,
\end{equation}
where $F_j$ and $U_{r_{j}}$ were the bell's applied contraction force and contraction velocity at the $j^{th}$ time step, and $V_{dim}$ was the temporally-averaged dimensional forward swimming speed during the $N$ time points considered. The term $\frac{1}{\rho f^2 D^3}$ non-dimensionalizes the data, providing a cost of transport metric normalized by the bell's driving frequency and size. $\rho$ is the fluid's density. Note that our dimensionless cost of transport defined above is similar to the energy-consumption coefficient discussed by Bale et al. 2014 \cite{Bale:2014}.

Thus for each duty cycle combination, Re, and $f$ considered, a time-averaged dimensionless forward swimming speed ($1/\mathrm{St}$) and cost of transport ($COT$) were calculated. We used this information to first construct colormap visualizations, which offered the benefit of qualitatively identifying duty cycle combinations that led to greater swimming performance for each $f$ and Re. Next we computed the percent enhancement in performance (and percent impedance) to assess the benefit (or drawback) of different duty cycle combinations, by comparing each to the case involving only using a single duty cycle throughout its entire simulation. Using this data we also explored performance landscapes of swimming speed versus $COT$ in order to compare all of the simulation data at once.

Finally, to probe into the physical mechanism that promotes enhancements or decay in performance from duty cycle combinations we also computed the non-dimensional fluid circulation, pressure, and vertical momentum all inside of the bell. The non-dimensional circulation in the bell was computed in the following manner
\begin{equation}
    \label{eq:circulation} \Gamma^n = \int_{\Omega} \frac{\omega^n}{fD^2} dA \approx \sum_{i,j} \frac{ \omega_{ij}^n}{fD^2} dxdy,
\end{equation}
where $\Omega_k$ is the area inside the \textit{left} side of the bell, $\omega^n = \nabla\times\textbf{u}^n$ is the fluid vorticity at time point $n$, and $f$ and $D$ are the jellyfish bell's actuation frequency and diameter, respectively. The overall sum of pressure (\textit{bell pressure}, $\mathscr{P}^n$) and vertical momentum inside the bell ($\mathscr{M}_y$) were calculated similarly, i.e.,
\begin{align}
    \label{eq:yMomentum} \mathscr{M}_y^n = \int_{\Omega} \frac{v^n}{fD^3}\ dA \approx \sum_{i,j} \frac{ v_{ij}^n}{fD^3}\ dxdy. \\
    \label{eq:Pressure} \mathscr{P}^n = \int_{\Omega} \frac{P^n}{fD^2}\ dA \approx \sum_{i,j} \frac{ P_{ij}^n}{fD^2}\ dxdy. 
\end{align}
We only computed these quantities on the \textit{left} side of the bell due to the emergent symmetry of the system. Therefore when bell circulation is positive, that there is net positive fluid vorticity inside, which may correspond to a possible counterclockwise vortex. To reduce numerical errors due to sampling the data near boundary, the area inside the bell, $\Omega_k$, was selected such that the grid cells used were 5 grid spatial widths from the jellyfish's bell.

%
%

%
%

\section{Results}
\label{results}

A total of twenty different batches of jellyfish simulations were performed, each with a specific Reynolds number ($\mathrm{Re}=37.5,75,150$, or $300$) and stroke frequency ($f=0.50, 0.75, 1.125, 1.5$, or $1.875\ \mathrm{Hz})$. In each batch combinations of two different duty cycle contraction fractions were simulated, i.e., $p_1,p_2\in\{0.05,0.10,0.15,\ldots,0.95\}$. This allowed us to both quantitatively and qualitatively determine duty cycle combinations that might enhance or impede swimming performance.

\begin{figure}[H]
    \centering
    \includegraphics[width=0.78\textwidth]{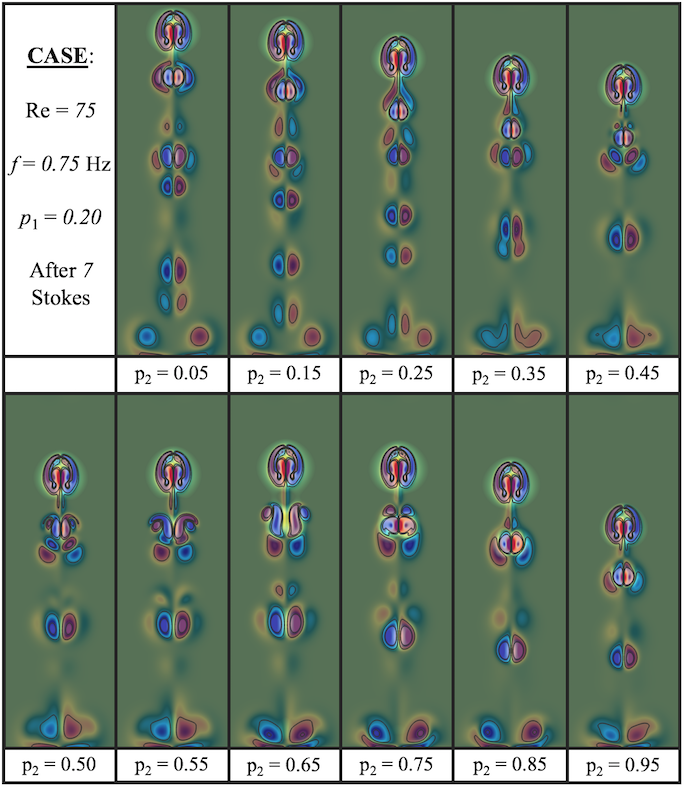}
    \caption{The positions and vortex wakes of various duty cycle jellyfish cases after 7 strokes for the case when $\mathrm{Re}=75$, $f=0.75\ \mathrm{Hz}$, and $p_1=0.20$.}
    \label{fig:Vorticity_Re75}
\end{figure}

\begin{figure}[H]
    \centering
    \includegraphics[width=0.95\textwidth]{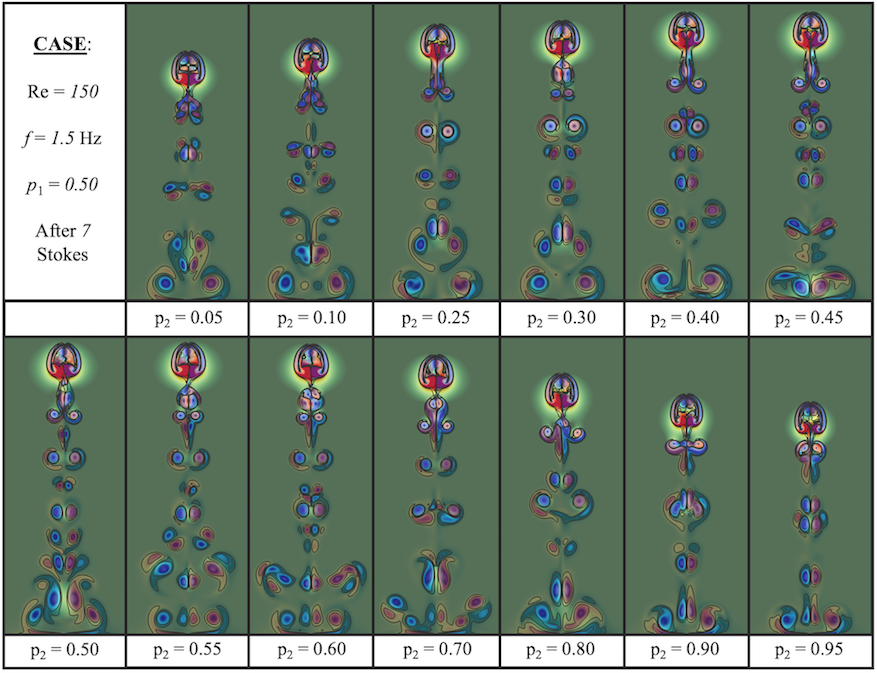}
    \caption{The positions and vortex wakes of various duty cycle jellyfish cases after 7 strokes for the case when $\mathrm{Re}=150$, $f=1.5\ \mathrm{Hz}$, and $p_1=0.50$.}
    \label{fig:Vorticity_Re150}
\end{figure}

Snapshots of a jellyfish's vortex wake are given in Figures \ref{fig:Vorticity_Re75} and \ref{fig:Vorticity_Re150} after the 7th stroke cycle. Figure \ref{fig:Vorticity_Re75} provides snapshots for the case when $\mathrm{Re}=75$, $f=0.75\ \mathrm{Hz}$, and $p_1=0.20$, for a variety of $p_2$ values. Figure \ref{fig:Vorticity_Re150} presents similar data but for $\mathrm{Re}=150$, $f=1.5\ \mathrm{Hz}$, and $p_1=0.50$. These figures illustrate three important ideas. The first is that there exist optimal combinations that lead to faster swimming, determined by the position of the jellyfish alone. Second, nonlinear relationships in performance exist for different duty cycle combinations. The third is that there are a number of combinations that seem to lead to the similar a distance swam after 7 stroke cycles. In the latter, the downstream vortex wakes are also topologically different. While their swimming speeds may be near identical, no information is present on the cost of transport through these visualizations. Without any additional temporal fluid dynamics information, individual vortex wake snapshots from a single moment in time may not be reliable, nor sufficient, indicators for interpreting swimming performance  \cite{Floryan:2020}. 

\begin{figure}[H]
    \centering
    \includegraphics[width=0.9\textwidth]{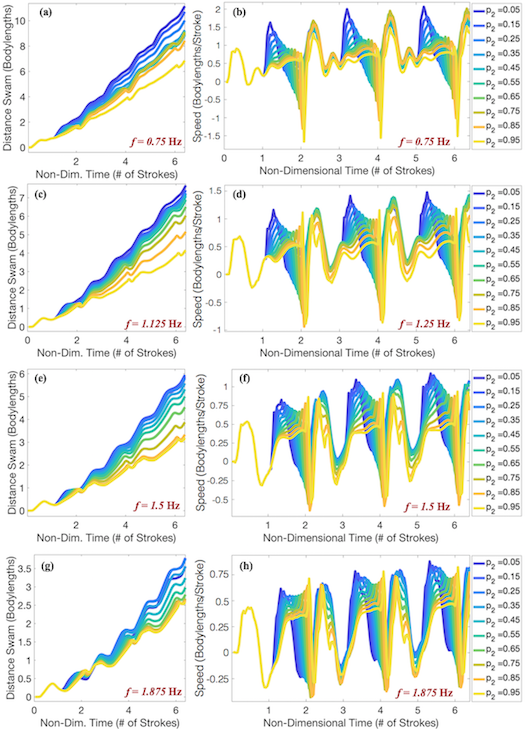}
    \caption{The distances swam (left column) and swimming speeds (right column) over time for cases with $\mathrm{Re}=75$, $p_1=0.20$, and a variety of $f$ and $p_2$, i.e., (a)-(b) $f=0.75\ \mathrm{Hz}$, (c)-(d) $f=1.125\ \mathrm{Hz}$, (e)-(f) $f=1.5\ \mathrm{Hz}$, and (g)-(h) $f=1.875\ \mathrm{Hz}$.}
    \label{fig:TimeEvo_Re75_VaryFreq}
\end{figure}

Temporal data is given in Figure \ref{fig:TimeEvo_Re75_VaryFreq} for the distance swam and swimming speed for the case of $\mathrm{Re}=75$, $p_1=0.20$, multiple stroke frequencies, and a variety of $p_2$ values. There is an inverse relationship between frequency and maximal distance swam; higher frequencies generally lead to impeded swimming \cite{Miles:2019}. Moreover, due to the combinations of two different duty cycles, the swimming speed waveform varies as $p_2$ changes, but the morphing appears continuous in $p_2$. As frequency changes, the waveform also changes. Interestingly, for the opposite case of a given frequency and multiple $\mathrm{Re}$, the waveform does not substantially vary in shape, only in amplitude, see Figure \ref{fig:TimeEvo_f0pt75_VaryRe} in the Supplemental Materials. This suggests that behavior across Re is conserved when combining different duty cycles.

\begin{figure}[H]
    \centering
    \includegraphics[width=0.9\textwidth]{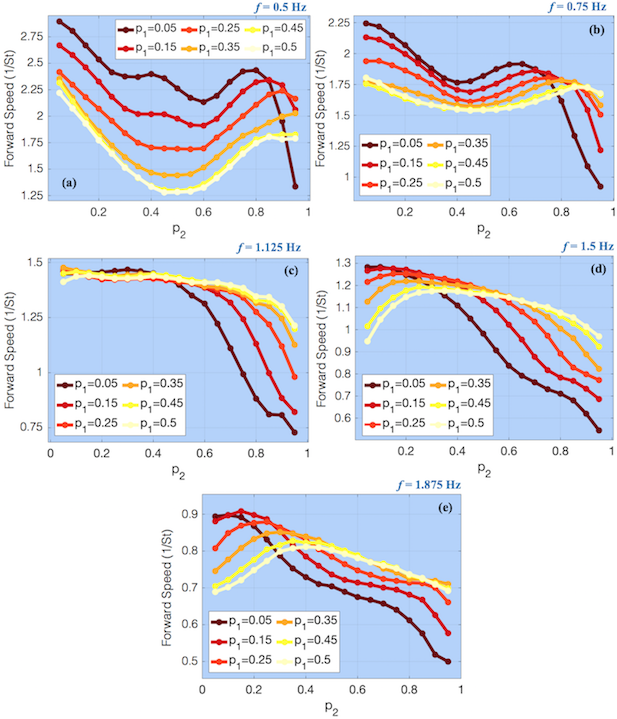}
    \caption{Plots of the time-averaged swimming speed against $p_2$ for various $p_1$ values when $\mathrm{Re}=75$ for (a) $f=0.50\ \mathrm{Hz}$, (b) $f=0.75\ \mathrm{Hz}$, (c) $f=1.125\ \mathrm{Hz}$, (d) $f=1.5\ \mathrm{Hz}$, and (e) $f=1.875\ \mathrm{Hz}$.}
    \label{fig:Plot_Speed_Re75_DiffFreq}
\end{figure}

Perhaps more informative than time evolution plots for performance comparisons, are the plots of the time-averaged swimming speeds given in Figure \ref{fig:Plot_Speed_Re75_DiffFreq}. Each data point shown in a particular plot is representative of the case when $\mathrm{Re}=75$ for a specific $f$, and for a different $p_1$ and $p_2$ combination. We note that the case of $\mathrm{Re}=75$, $f\sim0.75\ \mathrm{Hz}$ and $p_1=p_2=0.2$ roughly corresponds to that of a `large' \textit{Sarsia tubulosa} (see Table \ref{table:duty_cycles}) and that the corresponding range of average speeds associated with it in Figure \ref{fig:TimeEvo_Re75_VaryFreq} are consistent. As frequency varies across the figure as a whole, the nonlinear dependence on $p_2$ also changes, for a specific $p_1$. In some cases, relative maximal and minimal speeds emerge for different $p_2$, i.e., absolute maximal and minimal speeds are different than these values. In particular, maximal swimming speeds do \textit{not} always occur when $p_1=p_2$, i.e., benefits do exist when stitching together different duty cycle combinations. 

For example, consider the case in Figure \ref{fig:Plot_Speed_Re75_DiffFreq}b when $f=0.75\ \mathrm{Hz}$ and $p_1=0.05$. There is a relative minimum and maximum when $p_2=0.40$ and $0.60$, respectively. However, the absolute minimum and maximal speed occurs at the extremes in $p_2$, i.e., $p_2=0.95$ and $0.05$, respectively. On the other hand, the behavior is substantially different when $p_1=0.50$. When $p_1=0.50$, maximal speeds occur near the extreme values of $p_2$, while minimal speeds occur when $p_2$ is close to $p_1$, i.e., $p_2=0.50$. The case involving frequency $f=0.75\ \mathrm{Hz}$ (and assuming $p_1=p_2=0.50$) was previously found to be an optimal frequency which led to enhanced swimming speed due to the resonance properties of the bell-fluid system \cite{Hoover:2015}. These results indicate that combining different duty cycles could either enhance swimming speed (or impede it) beyond previously reported optimal performance parameters.

\begin{figure}[H]
    \centering
    \includegraphics[width=0.99\textwidth]{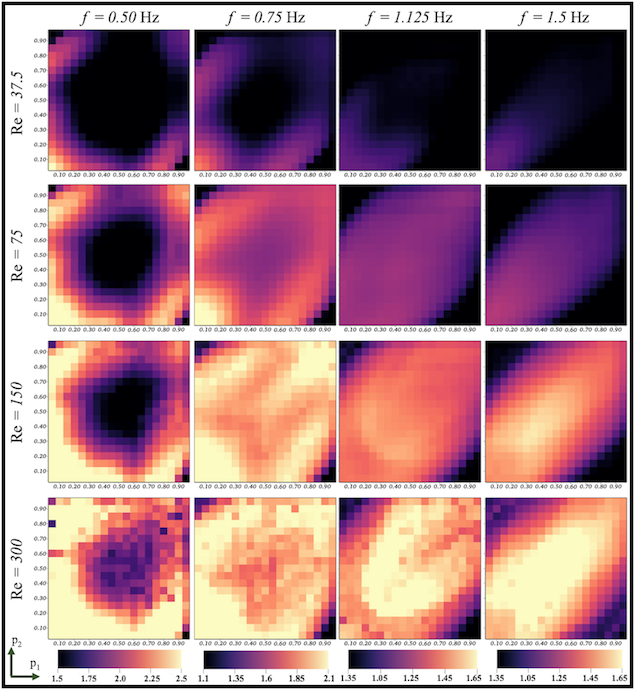}
    \caption{Colormaps representing the time-averaged dimensionless swimming speed (bodylengths/stroke) for different duty cycle combinations for each $\mathrm{Re}$ and $f=0.50,0.75,1.125$, and $1.5$ Hz. The case when $f=1.875\ \mathrm{Hz}$ is provided in Figure \ref{fig:Colormaps_f1pt875}.}
    \label{fig:Colormaps_Speed}
\end{figure}

Optimal combinations of duty cycles $p_1$ and $p_2$ for swimming speed emerge within the colormaps shown in Figure \ref{fig:Colormaps_Speed}. The horizontal and vertical axis in each colormap represent $p_1$ and $p_2$ values, respectively. Data is provided for the cases involving $Re=37.5, 75, 150$, and $300$ for $f=0.50,0.75,1.125$, and $1.5\ \mathrm{Hz}$. The cases involving $f=1.875\ \mathrm{Hz}$ are given in Figure \ref{fig:Colormaps_f1pt875} in the Supplemental Materials. As expected each colormap appears symmetric over the forward diagonal suggesting that a combination of $p_1$ then $p_2$ led to the same swimming speed as $p_2$ then $p_1$. For each specific colormap of a given $\mathrm{Re}$ and $f$, optimal duty cycle combinations emerge for higher swimming speeds. Furthermore, such combinations appear consistent given a frequency, $f$, across all of the $\mathrm{Re}$ cases simulated. That is, the optimal combinations for one $\mathrm{Re}$ appear to have the same pattern for a different $\mathrm{Re}$, given the same frequency. Furthermore, as frequency increases (from left to right) exhibiting a uniform duty cycle ($p_1=p_2$) appears more favorable than using two different. This was also observed in Figure \ref{fig:Plot_Speed_Re75_DiffFreq}.\\

\begin{figure}[H]
    \centering
    \includegraphics[width=0.99\textwidth]{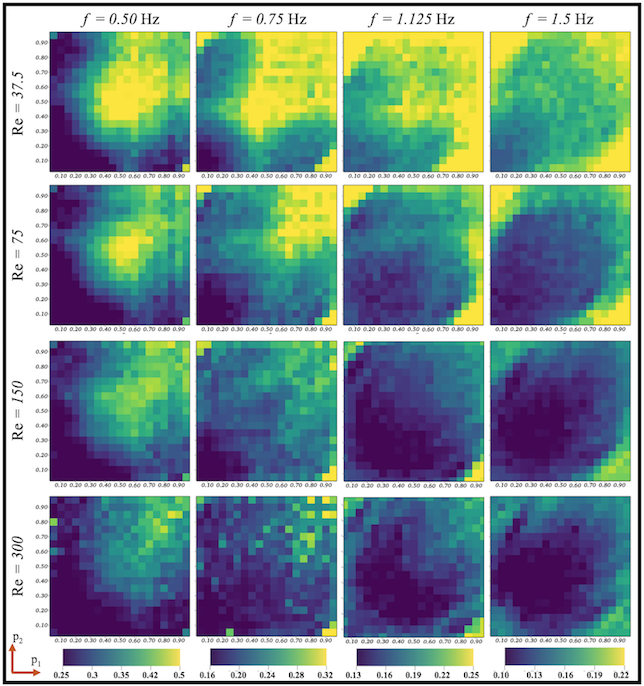}
    \caption{Colormaps representing the time-averaged dimensionless cost of transport for different duty cycle combinations for each $\mathrm{Re}$ and $f=0.50,0.75,1.125$, and $1.5$ Hz. The case when $f=1.875\ \mathrm{Hz}$ is provided in Figure \ref{fig:Colormaps_f1pt875}.}
    \label{fig:Colormaps_COT}
\end{figure}

The cost of transport data is provided in a similar fashion in Figure \ref{fig:Colormaps_COT} to complement the swimming speed data of Figure \ref{fig:Colormaps_Speed}. As expected each COT colormap panel is approximately symmetric over the forward diagonal. In each colormap, regions of lower COT emerge. Generally lower $p_1$ and $p_2$ duty cycle combinations lead to lower COT across each $\mathrm{Re}$ and $f$ panel. Similarly, duty cycle combinations involving either higher $p_1$ and $p_2$ values, or a low $p_1$ and high $p_2$ (or vice-versa) lead to higher COT. However, as $f$ increases, the lower COT region grows to encompass more $p_1$ and $p_2$ combinations. Interestingly, lower COT branches emerge as $f$ also increases, e.g., in the $\mathrm{Re}=300$ and $f=1.5\ \mathrm{Hz}$ or $f=1.875\ \mathrm{Hz}$ panels (see Figure \ref{fig:Colormaps_f1pt875} in the Supplemental Materials). Overall, we see a similar pattern with the swimming speed data in which repetitive duty cycles ($p_1=p_2$) are not always optimal across different $\mathrm{Re}$ and $f$ in minimizing COT. \\

To better assess the enhancement in swimming performance, we computed the relative change percentage in performance for each $(p_1,p_2)$ combination against its associated repetitive $(p_1,p_1)$ case. An illustration of which simulations are being compared is given in Supplemental Figure \ref{fig:Relative_Change_Idea} . The percent change was computed in the follow manner 
\begin{equation}
    \left(\% \mbox{\ change } \phi \right)_j = \frac{ \phi_{j,k}-\phi_{j,j} }{ \phi_{j,j} }\times100,
\end{equation}
where $\phi$ is a particular performance metric and $phi_{j,k}$ represents that metric's value for a specific combination of $p_1$ and $p_2$. For boosts in performance, the relative change would be positive for speed but negative for cost of transport. That is, a negative relative change for speed suggests an impedance in swimming speed, but a negative relative change for the cost of transport suggests an increase in efficiency, as cost of transport would have decreased. Supplemental Figures \ref{fig:Colormaps_Change_Speed} and \ref{fig:Colormaps_Change_COT} provide colormaps of the regions of enhancement or impedance in the swimming speed and cost of transport data.

\begin{figure}[H]
    \centering
    \includegraphics[width=0.85\textwidth]{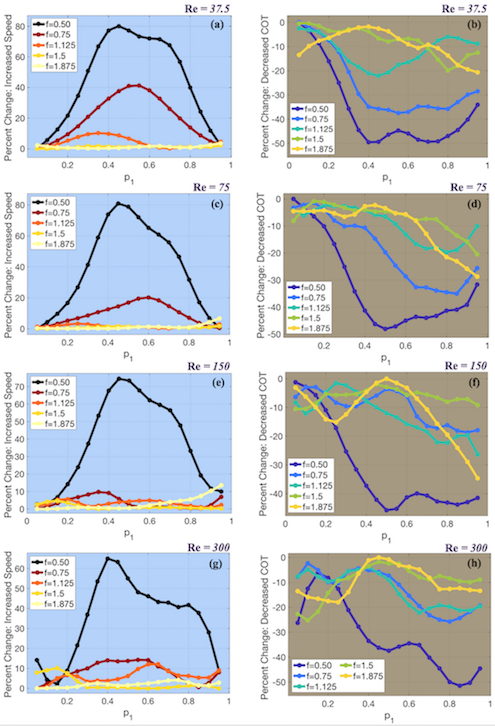}
    \caption{The maximal enhancement in swimming speed (left column) and cost of transport (right column) from different duty cycles combinations is provided. For a given $p_1$, the corresponding optimal $p_2$ was selected that positively affected performance for each case. The data is divided by $\mathrm{Re}$: (a)-(b) $\mathrm{Re}=37.5$, (c)-(d) $\mathrm{Re}=75$, (e)-(f) $\mathrm{Re}=150$, and (g)-(h) $\mathrm{Re}=300$ and every $f$ is given in each plot.}
    \label{fig:Percent_Change_Enhance}
\end{figure}

Figure \ref{fig:Percent_Change_Enhance} provides the maximal relative percent change for swimming speed (left side) and cost of transport (right side), for duty cycle combinations $(p_1,p_2)$ against a specific repetitive $(p_1,p_1)$ case. Thus, Figure \ref{fig:Percent_Change_Enhance} provides the maximum positive and negative values for enhancements in speed and COT, respectively, from each column of Figures \ref{fig:Colormaps_Change_Speed} and \ref{fig:Colormaps_Change_COT}. The most substantial percent increases in speed were observed in the lowest frequency case ($f=0.50\ \mathrm{Hz}$) for every Re considered. These percent increases were on the order of $\sim60-80\%$ for cases involving $p_1=0.4, 0.45$, or $0.50$. Similarly, maximal benefits were observed in COT when $f=0.50\ \mathrm{Hz}$, in every Re case. These maximal reductions in COT were on the order of $50\%$. While the cases involving either $\mathrm{Re}=37.5$ or $75$ and $f=1.5$ or $1.875\ \mathrm{Hz}$ showed only minimal benefits to combining different duty cycles in terms of speed, reductions in COT of up to $15-20\%$ were possible for particular duty cycle combinations. Similar observations were seen for other higher $f$ and Re cases. Nonlinear dependence on $p_1$ emerges for the maximal percent increase/decrease across each $\mathrm{Re}$ and $f$ considered. Moreover, the nonlinear behavior exhibited is not the same between the relative percent changes in speed and COT. 

\begin{figure}[H]
    \centering
    \includegraphics[width=0.85\textwidth]{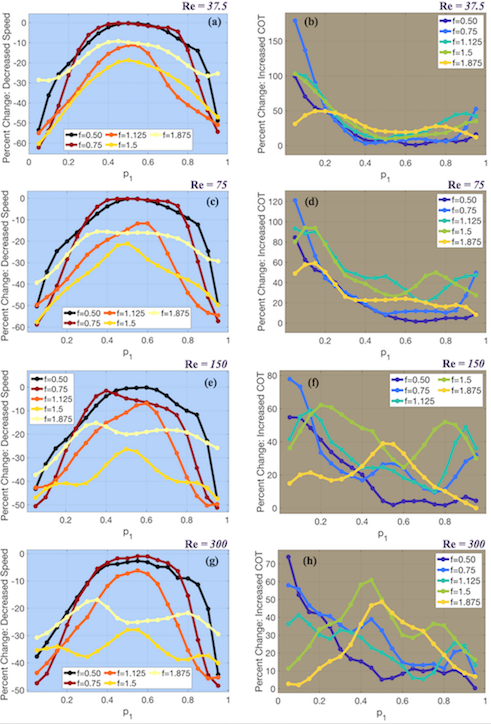}
    \caption{The maximal impedance in swimming speed (left column) and cost of transport (right column) from different duty cycles combinations is provided. For a given $p_1$, the corresponding worse $p_2$ was selected that negatively impacted performance for each case. The data is divided by $\mathrm{Re}$: (a)-(b) $\mathrm{Re}=37.5$, (c)-(d) $\mathrm{Re}=75$, (e)-(f) $\mathrm{Re}=150$, and (g)-(h) $\mathrm{Re}=300$ and every $f$ is given in each plot.}
    \label{fig:Percent_Change_Impede}
\end{figure}

Unfortunately, there are dire consequences to choosing duty cycle combinations poorly, as Figure \ref{fig:Percent_Change_Impede} illustrates. Figure \ref{fig:Percent_Change_Impede} shows the opposite of Figure \ref{fig:Percent_Change_Enhance}. That is, it presents the maximal \textit{impedance} in performance from different duty cycle combinations - the maximal decrease in speed and maximal increase in cost of transport. As suggested previously in Figure \ref{fig:Colormaps_Speed}, duty cycle combinations involving extremal contraction fractions $p_1$ or $p_2$ can result in substantially decreased swimming speeds, roughly $\gtrsim30\%$ for each $\mathrm{Re}$ and $f$. Moreover, poor combinations could lead to percent increases in COT greater than $100\%$, e.g., $\mathrm{Re}=37.5$ and lower $p_1$. It appears that greater reductions in speed are possible for extremal $p_1$ values. Furthermore, many worse-case possibilities led to greater than $10\%$ decreases in speed when $p_1\lesssim0.25$ or $p_1\gtrsim0.75$. There were many worse-case combinations that led to greater than $20\%$ increases in COT; however, there was more non-linear fluctuation in which $p_1$ values these cases were associated with. In comparison, only a few duty cycle combinations led to increases and decreases of $10\%$ and $20\%$ or more in speed and COT, respectively, when $f>0.5\ \mathrm{Hz}$ (Figure \ref{fig:Percent_Change_Enhance}).

%
%
\subsection{Physical mechanisms and fluid dynamics analysis}
\label{results:Physical_Analysis}

Before discussing the physical mechanisms that lead to increases in swimming speed, we will briefly describe some of the key observable features in a standard uniform duty cycle case's temporal waveforms. We will focus our efforts on the lower frequency cases, particularly when $f=0.5\ \mathrm{Hz}$; this is when effects of combining duty cycles were most substantially observed. Since Figure \ref{fig:TimeEvo_f0pt75_VaryRe} suggested similar behavior across all $\mathrm{Re}$ considered, we focused our efforts on cases involving $\mathrm{Re}=37.5$. 

\subsubsection{Temporal data analysis and passive energy recapture}

\begin{figure}[H]
    \centering
    \includegraphics[width=0.925\textwidth]{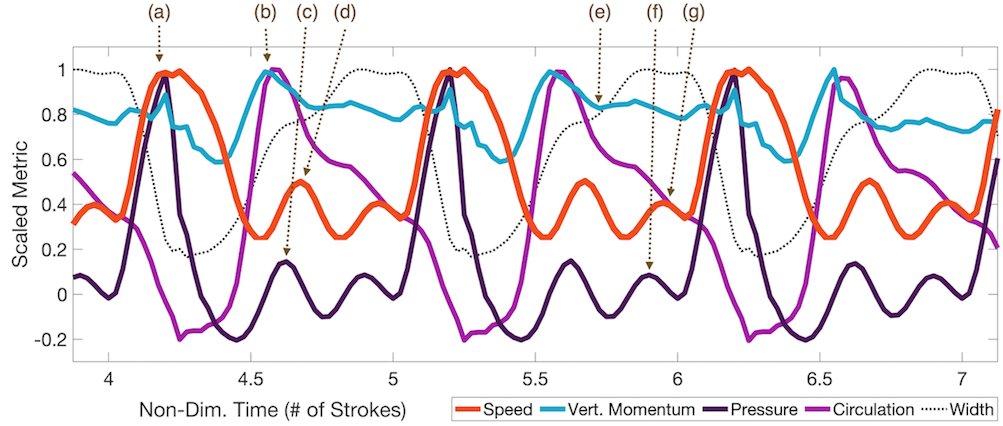}
    \caption{Time evolution of the dimensionless swimming speed, vertical momentum in the bell, bell pressure, bell circulation, and bell width, each scaled by their respective maximum, i.e., for data $q$, $q/\max|q|$ is plotted, when $f=0.5\ \mathrm{Hz}$, $\mathrm{Re}=37.5$, and $(p_1,p_2)=(0.2,0.2)$. (a) First peak in swimming speed, vertical momentum, and pressure due to bell contraction (width decreases). As fluid is expelled from the bell, circulation decreases. (b) As bell expands (width increases), the stopping vortex begins to form and is pulled into the bell, which corresponds to an increase is bell circulation. As circulation increases, the fluid's vertical momentum in the bell also increases and peaks nearly when the circulation peaks. (c) Shortly after bell circulation and vertical momentum increase, bell pressure peaks followed by (d) a second peak in swimming speed. Processes (b)-(d) are referred to as \textit{passive energy recapture} (PER). (e) Bell expansion continues (after a brief pause) and both circulation and vertical momentum stay nearly constant for a short time. This provides conditions for (f) bell pressure to once again peak and (g) swimming speed to peak shortly thereafter. This emergent third peak can also be attributed to PER. 
}
\label{fig:p1p2_20}
\end{figure}

Figure \ref{fig:p1p2_20} provides temporal data that illustrates when \textit{passive energy recapture} (PER) occurs \cite{Gemmell:2013,Gemmell:2018}. First, notice that when the bell contracts, its width decreases alongside bell circulation. Swimming speed also increases at this time and then begins to quickly decrease once the bell begins its expansion phase. This is due to the formation of a stopping vortex under the bell during the onset of expansion (see Supplemental Figure \ref{fig:Vortex_Defn} for vortex definitions). The stopping vortex gets pulled into the bell as expansion continues, resulting in an increase in bell circulation, followed by an increase in bell pressure. The increase is pressure provides additional upwards thrust through suction \cite{Gemmell:2015}. This overall process results in an additional boost in swimming speed, i.e., PER. 

Notably, Figure \ref{fig:p1p2_20} shows that in this case that there are two PER-related swimming speed peaks, which is thought to be possible due to bell expansion pausing and then restarting during the expansion phase. To the authors' knowledge this second peak has not been observed in real jellyfish. Similar PER processes were observed for cases involving $f\lesssim 0.75\ \mathrm{Hz}$ when $p_1=p_2\lesssim0.5$, for all $\mathrm{Re}$ studied here \cite{Miles:2019}. 

Boosts in swimming performance from stitching multiple duty cycles together can be observed by similar data analysis. Combining different duty cycles does not simply connect waveforms, like those shown in Figure \ref{fig:p1p2_20}. Instead the fluid dynamic benefits from cycle-to-cycle are observed in the temporal data, suggesting why certain combinations lead to enhanced performance. We will review two cases in which boosts in swimming speed are observed - when opposite duty cycles are combined, e.g., $(p_1,p_2)=(0.2,0.8)$, and when either a low or high duty cycle is combined with a mid-range duty cycle of $0.50$, $(p_1,p_2)=(0.5,0.2)$ and $(p_1,p_2)=(0.5,0.8)$. As a reference, the dimensionless swimming speeds (\textit{averaged across only the $5^{th}$ and $6^{th}$ actuation cycles}) for the cases discussed in the proceeding analysis are given in Table \ref{table:data_analysis_speeds}.

\begin{table}[h!]
\centering
\begin{tabular}{ |c|c|c| } 
 \hline
  \multicolumn{3}{|c|}{Swimming Speeds} \\ \hline
 & Duty Cycle Combination    & Swimming Speed ($1/\mathrm{St}$) \\ \hline 
  \multirow{3}{*}{$\substack{\text{low-high} \\ \text{combinations}}$} & (0.2,0.2)  & 1.77   \\ 
 &(0.2,0.8)  & 1.64 \\ 
 &(0.8,0.8)  & 1.28  \\ \hline \hline
 \multirow{3}{*}{$\substack{\text{mid-range} \\ \text{combinations}}$} & (0.5,0.5)  & 0.95  \\ 
 &(0.5,0.2)  & 1.43   \\ 
 &(0.5,0.8)  & 1.17   \\ \hline
\end{tabular}
\caption{Time-averaged dimensionless swimming speeds across the $5^{th}$ and $6^{th}$ actuation cycles, rather than the $7^{th}$ and $8^{th}$, to complement the temporal analysis below.}
\label{table:data_analysis_speeds}
\end{table}

\begin{figure}[H]
    \centering
    \includegraphics[width=0.95\textwidth]{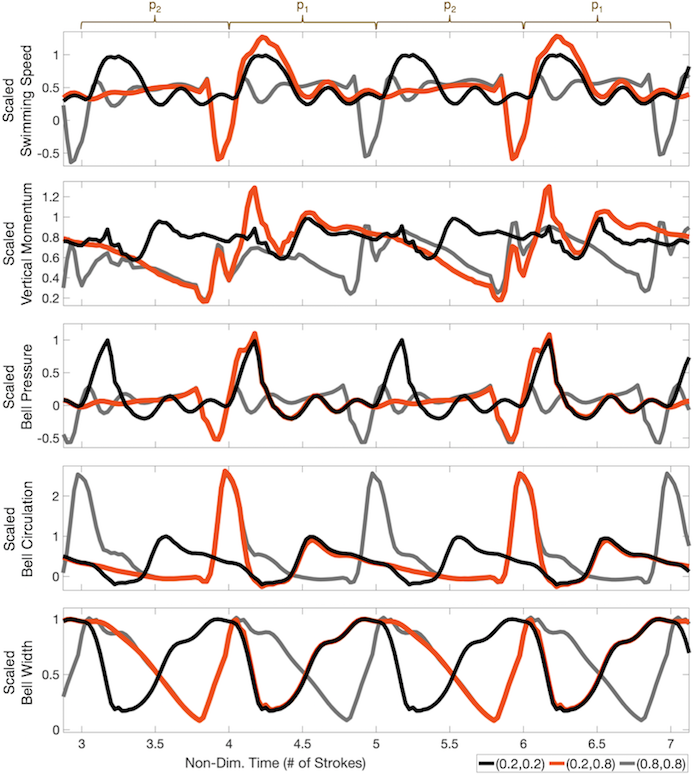}
    \caption{Time evolution of the dimensionless swimming speed, vertical momentum in the bell, bell pressure, bell circulation, and bell width, each scaled by the maximum for the \textit{uniform} duty cycle case, $q_u$, i.e., for data $q$, $q/\max|q_u|$ is plotted, when $f=0.5\ \mathrm{Hz}$, $\mathrm{Re}=37.5$, and three different duty cycle combinations: $(p_1,p_2)=\{(0.2,0.2),(0.2,0.8),(0.8,0.8)\}$. }
\label{fig:p1p2_opposites}
\end{figure}

For a combination of a high and low duty cycle, the swimming speed waveform shows two interesting dynamics, see Figure \ref{fig:p1p2_opposites} for temporal data from the case when $f=0.5\ \mathrm{Hz}$ and $\mathrm{Re}=37.5$ for 3 duty cycle combinations: low-low, low-high, and high-high. Note that the low-high case is highlighted, i.e., $(p_1,p_2)=(0.2,0.8)$. First, when transitioning from a high-to-low duty cycle  ($0.8\rightarrow0.2$), the peak in swimming speed is larger than that of the uniform duty cycle case ($p_1=p_2=0.2$). Similarly, so are the speed peaks associated with PER. For the case shown, the speed's peak amplitude is $\sim25\%$ higher. This may be the result of having a high bell circulation at the end of a fast expansion phase (from a high duty cycle case), followed by a quick bell contraction phase (from a low duty cycle case). Due to a fast expansion phase, the bell is filled with higher energetic fluid, which then gets expelled from during a quick contraction phase, thus, creating a more energetic jet, i.e., higher momentum injection into the wake \cite{Costello:2020}. Also, the bell's internal pressure increases at a faster rate compared to the uniform case, which results in a larger peak vertical momentum in the bell. These dynamics are likely products of vortex-vortex interactions from cycle-to-cycle, see Section \ref{results:vortex-vortex}, and also jellyfish virtual wall effects \cite{Gemmell:2021}. However, before that boost in speed occurs, at the end of the previous cycle, the bell quickly decelerates and is even pulled downward during the fast expansion phase. This is due to the fast formation of a powerful stopping vortex, giving rise to high circulation in the bell. This corresponds to when the bell pressure reaches its minimal, negative value, suggesting a negative suction downwards during this time. 

Second, while a large negative speed peak emerges at the end of the high duty cycle portion in the $(p_1,p_2)=(0.2,0.8)$, the speed waveform shows that the low-high case is able to offset some of this loss. We will continuing comparing this case to the uniform $p_1=p_2=0.2$ case. This offset is achieved by a more steady swimming speed while the bell is slowly contracting. This speed is higher than the first PER peak of the uniform duty cycle case throughout its elongated contraction phase. Thus, while the speed in uniform duty cycle case oscillates during this time, the high duty cycle's speed does not. The steadier speed in the high duty cycle case corresponds to steadier bell pressure and a slowly dissipating vertical momentum in the bell as fluid is expelled. 

\begin{figure}[H]
    \centering
    \includegraphics[width=0.99\textwidth]{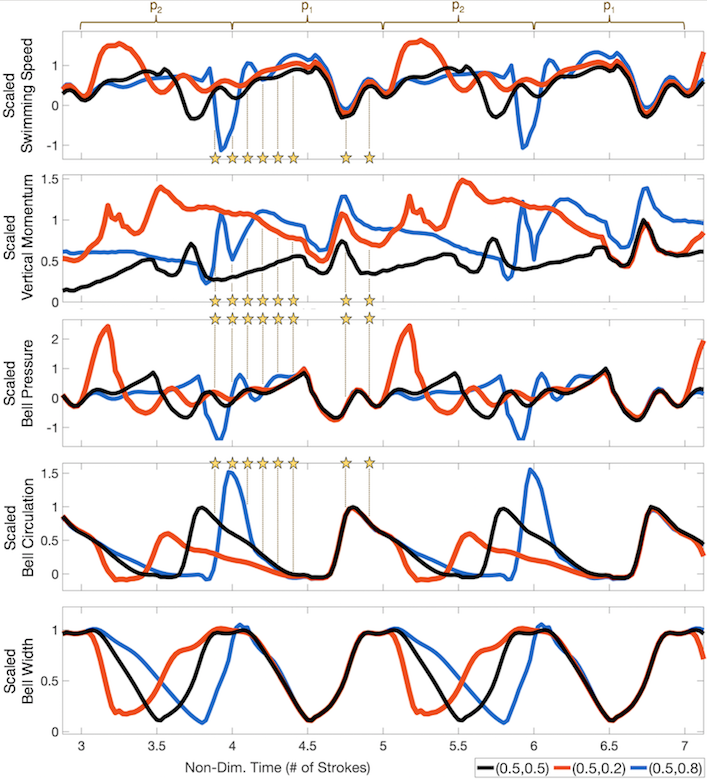}
    \caption{Time evolution of dimensionless swimming speed, vertical momentum in the bell, bell pressure, bell circulation, and bell width, each scaled by the maximum for the uniform duty cycle case, $q_u$, i.e., for data $q$, $q/\max|q_u|$ is plotted, when $f=0.5\ \mathrm{Hz}$, $\mathrm{Re}=37.5$, and three different duty cycle combinations: $(p_1,p_2)=\{(0.5,0.2),(0.5,0.5),(0.5,0.8)\}$. The stars correspond to visualized time-points of Figure \ref{fig:p1p2_0pt5_Vorticity_Pressure}}
\label{fig:p1p2_middle}
\end{figure}

For combinations involving a mid-range duty cycle and either a low or high duty cycle, e.g., $(0.5,0.2)$ or $(0.5,0.8)$, respectively, similar elements of the low-high duty cycle combination case discussed above are observed. For example, when $(p_1,p_2)=(0.5,0.2)$, during the $0.2$ duty cycle portion, maximal peaks in swimming speed again arise - the first due to bell contraction and two that follow due to PER effects, see Figure \ref{fig:p1p2_middle}. It is also worth mentioning that the uniform duty cycle case $(p_1,p_2)=(0.5,0.5)$ only emits one PER peak. Moreover, the swimming speed is greater in the $(0.5,0.2)$ case compared the uniform $(0.5,0.5)$ case, throughout its entire cycle but a short time (roughly $15\%$ of the entire cycle) before its first PER peak. During the next cycle when they both have a common duty cycle of $0.5$, the $(0.5,0.2)$ case's swimming speed is slightly greater than the uniform case throughout. This could be the result of both its bell starting with a higher vertical momentum as well as maintaining a higher level throughout, as their bell pressures are similar throughout this common portion. However, entering this common cycle their bell circulations are also substantially different. This suggests that cycle-to-cycle fluid dynamics in the bell and vortex interactions may be responsible for these differences in speed. 

On the other hand, the case when $(p_1,p_2)=(0.5,0.8)$ shows some elements of the low-high $(0.2,0.8)$ case from Figure \ref{fig:p1p2_opposites}. That is, during the high duty cycle portion $(p_2=0.8)$, swimming speed is steadier until the quick bell expansion phase commences. Although, a notable difference is that during the middle duty cycle portion $(p_1=0.5)$, two speed peaks emerge due to bell contraction. Both speed peaks also correspond to peaks in bell pressure. This is different than the $(0.2,0.8)$ case, when only one such peak emerged due to the bell's contraction. Moreover, in the $(0.5,0.8)$ case, swimming speed is greater than or equal to that of the uniform $p_1=p_2=0.5$ case during the $p_1$ portion of the cycle. When the speed reaches its maximum instantaneous value in the uniform $(0.5,0.5)$ duty cycle case, the speed of the $(0.5,0.8)$ case is $\sim30\%$ greater. However, the $(0.5,0.8)$ case, still elicits a rapid acceleration towards negative swimming speeds at the end of the high duty cycle portion ($p_2=0.8$). This again corresponds to a minimal, negative bell pressure, suggesting the bell gets suctioned downwards because of quick expansion kinematics. In fact, this process occurs for all cases presented in Figure \ref{fig:p1p2_middle} when their common duty cycle of $0.5$ is used. That is, a duty cycle of $0.5$ still leads to the formation of a powerful enough stopping vortex, due to a shorter expansion phase time-scale, such that the bell quickly decelerates and gets suctioned downward by a low, negative overall bell pressure. Although, as the formation time of the stopping vortex was shorter in the $0.8$ duty cycle case than the $0.5$ case, this negatively interfering suction process has more profound effect in decreasing swimming speeds in the $0.8$ case than the $0.5$.

\subsubsection{Vortex-vortex dynamics}
\label{results:vortex-vortex}

\begin{figure}[H]
    \centering
    \includegraphics[width=0.95\textwidth]{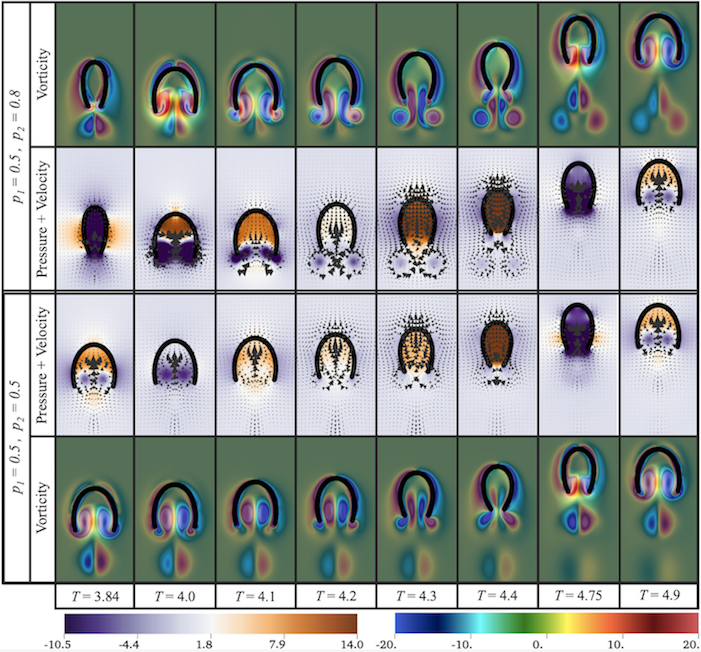}
    \caption{Snapshots of dimensionless vorticity and pressure along with the velocity field field for duty cycle combinations $(p_1,p_2)=(0.5,0.5)$ and $(p_1,p_2)=(0.5,0.8)$ when $\mathrm{Re}=37.5$ and $f=0.5\ \mathrm{Hz}$, during the $5^{th}$ actuation cycle when they both have a common duty cycle ($p_1=0.50$). The time-points shown here correspond to the stars in Figure \ref{fig:p1p2_middle}.}
\label{fig:p1p2_0pt5_Vorticity_Pressure}
\end{figure}

Besides the bell kinematics alone, vortex-vortex interactions play an important role in enhancing performance from cycle-to-cycle. Consider the two scenarios discussed above where we included a high duty cycle in the combination, i.e., $(0.2,0.8)$ and $(0.5,0.8)$. We observed boosts in the instantaneous swimming speed beyond what was to be expected from their respective uniform duty cycle case, $(0.2,0.2)$ or $(0.5,0.5)$. These boosts occurred when going from the higher-to-lower duty cycle kinematics, either $0.8\rightarrow0.2$ or $0.8\rightarrow0.5$ (see the first speed peaks shortly after $T=4$ in Figures \ref{fig:p1p2_opposites} and \ref{fig:p1p2_middle}). Snapshots of the instantaneous vorticity, pressure and velocity fields are provided in Figure \ref{fig:p1p2_0pt5_Vorticity_Pressure} in order to qualitatively compare the $(0.5,0.8)$ and $(0.5,0.5)$ cases during the $5^{th}$ actuation cycle ($T$ is number of actuation cycles already performed). They both actuate their bells with a $0.5$ duty cycle in this cycle. At the beginning of this cycle ($T=4$), both jellyfish bells have the same shape (width, fineness); however, the internal and external fluid dynamics of the jellyfish differ. 

For example, in the $(0.5,0.8)$ case, the previous cycle's starting and stopping vortices still appear to be strongly interacting near the bottom of the bell. In fact, perhaps it is from these interactions that strong oppositely vortices develop just below the below the bell ($T=4.1$), that pull the previous cycle's stopping vortices downwards ($T\in[4.2,4.4]$). These stopping vortices are pulled out of its bell much sooner than the uniform $(0.5,0.5)$ case expels its previous cycle's stopping vortex (compare panels $T=4.2$ and $T=4.3$). When the $(0.5,0.8)$ case shows developed starting vortices (just under the tips of the bell) at $T=4.4$, they are still strongly interacting with a distinct stopping vortex from its previous cycle. By the end of the cycle, the vortex wakes are substantially different, which coincide with subtle qualitative differences in the wake's velocity field, too. Also at this point, the bell's internal dynamics appear qualitatively the same between the two cases, i.e., the bell's pressure, vorticity, and velocity field appear identical. Lastly, the vortex-vortex interactions during this cycle may also manifest themselves as differences in the pressure dynamics throughout the cycle, as in Figure \ref{fig:p1p2_0pt5_Vorticity_Pressure}. As Figures \ref{fig:p1p2_opposites} and \ref{fig:p1p2_middle} suggest, swimming speed follows the bell's internal pressure dynamics closely, i.e., speed increases or decreases (following short delays) and peaks shortly after pressure peaks.

To decipher the contribution from vortex-vortex interactions, swimming distances were computed across cycles 3, 5, and 7 when each pair, either $(0.2,0.2)$ and $(0.2,0.8)$, or $(0.5,0.5)$ and $(0.5,0.8)$, exhibited the same duty cycle, i.e., the same bell kinematics. Note that cycles 3,5, and 7 are all preceded by a higher duty cycle, either $0.8$ or $0.5$, respectively. These distances swam were then directly compared to the case of when the jellyfish was starting from rest, in a quiescent ambient fluid \cite{Gemmell:2021}, using the same duty cycle, either $0.2$ or $0.5$. Thus we compared distances swam in actuation cycles 3,5, and 7 to the first. Table \ref{table:distances_Swam} provides the distances swam in each single cycle for every case considered, while Table \ref{table:distances_percent_increase} provides complementary data giving the percent increase in distance traveled. Across all cases, there was a substantial increase in the distance swam (and hence also average swimming speed) during each individual cycle compared to the case when starting from rest. 

\begin{figure}[H]
    \centering
    \includegraphics[width=0.95\textwidth]{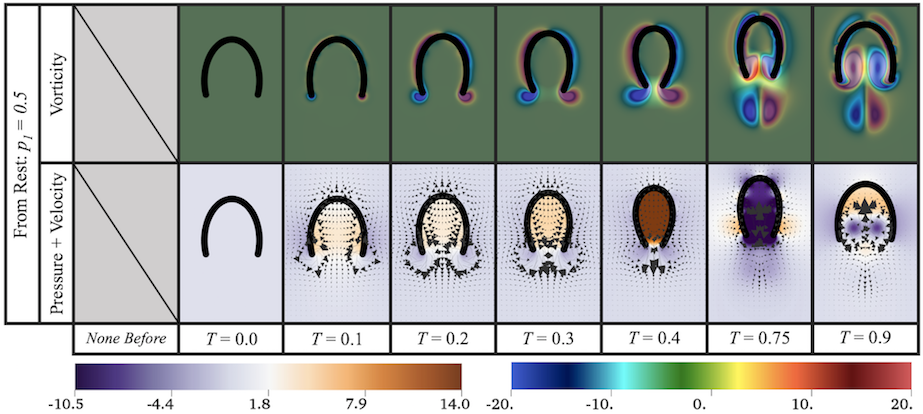}
    \caption{Snapshots of dimensionless vorticity and pressure along with the velocity field field for a jellyfish starting from rest in quiescent fluid, with a duty cycle of $0.5$ during its first actuation cycle when $\mathrm{Re}=37.5$ and $f=0.5\ \mathrm{Hz}$. The time-points shown here correspond to same proportion of time into an actuation cycle as Figure \ref{fig:p1p2_0pt5_Vorticity_Pressure}, or the stars in Figure \ref{fig:p1p2_middle}.}
\label{fig:p1p2_0pt5_Vorticity_Pressure_Rest}
\end{figure}

Two things are identified in this data. First, the coupling of fluid motion from cycle-to-cycle can benefit a jellyfish's locomotion performance, even when it performs the same kinematics. When the jellyfish starts from rest, the fluid is void of any ambient motion. Hence vortices persisting at the start of a new cycle, like those in Figure \ref{fig:p1p2_0pt5_Vorticity_Pressure}, must play an important role in cycle-to-cycle performance. Stopping vortices present in the bell at the onset of the next cycle may give rise to \textit{virtual wall effects}, as suggested by Gemmell et al. 2021 \cite{Gemmell:2021}. Furthermore, Figure \ref{fig:p1p2_0pt5_Vorticity_Pressure_Rest} explicitly shows that the vortex and pressure dynamics are qualitatively different when starting from rest. It isn't until roughly $75\%$ through its actuation cycle before the dynamics begin to visually resemble the uniform $(0.5,0.5)$ case from Figure \ref{fig:p1p2_0pt5_Vorticity_Pressure}. Second, the data illustrates that the vortex dynamics emerging from non-uniform duty cycle combinations, like $(0.2,0.8)$ or $(0.5,0.8)$ can provide additional performance boosts when compared to the case starting at rest and their uniform duty cycle counterparts. The distances swam (and thereby their percent increases) involving the non-uniform duty cycle cases are greater than that of their respective uniform case (see Tables \ref{table:distances_Swam} and \ref{table:distances_percent_increase}). We note that these data does not contradict Table \ref{table:data_analysis_speeds} as it does not take into account the distance swam during the non-common duty cycle. 

\begin{table}[h!]
\centering
\begin{tabular}{ |c||c|c||c|c| } 
 \hline
 & \multicolumn{4}{|c|}{Distance Swam for Duty Cycle Combination} \\ \hline
 Cycle & (0.2,0.2) & (0.2,0.8) & (0.5,0.5) & (0.5,0.8) \\ \hline \hline
 1 (from rest) & 0.85 & 0.85  &   0.29     & 0.29  \\ \hline
 3             & 1.63 & 1.78  &   0.70     & 0.92\\ \hline
 5             & 1.77 & 2.01  &   0.94     & 1.31  \\ \hline
 7             & 1.79 & 2.04  &   0.97     & 1.33  \\ \hline
\end{tabular}
\caption{Distances swam (in bodylengths) across cycles 1,3,5, and 7, when the jellyfish exhibits the same duty cycle for comparison.}
\label{table:distances_Swam}
\end{table}

\begin{table}[h!]
\centering
\begin{tabular}{ |c||c|c||c|c| } 
 \hline
  & \multicolumn{4}{|c|}{Percent Increase for Duty Cycle Combination} \\ \hline
 Cycle (compared to 1) & (0.2,0.2) & (0.2,0.8) & (0.5,0.5) & (0.5,0.8) \\ \hline \hline
 3             & 91.8  & 109.4   &   141.4  & 217.2\\ \hline
 5             & 108.2 & 137.7   &  224.1   & 351.7  \\ \hline
 7             & 110.6  & 140.0  &  234.5   & 358.6  \\ \hline
\end{tabular}
\caption{The percent increase in distances swam across cycles 3,5, and 7 when compared to the distance swam during cycle 1.}
\label{table:distances_percent_increase}
\end{table}

%
%

\subsection{Performance Space Analysis}
\label{results:Performance_Spaces}

\begin{figure}[H]
    \centering
    \includegraphics[width=0.875\textwidth]{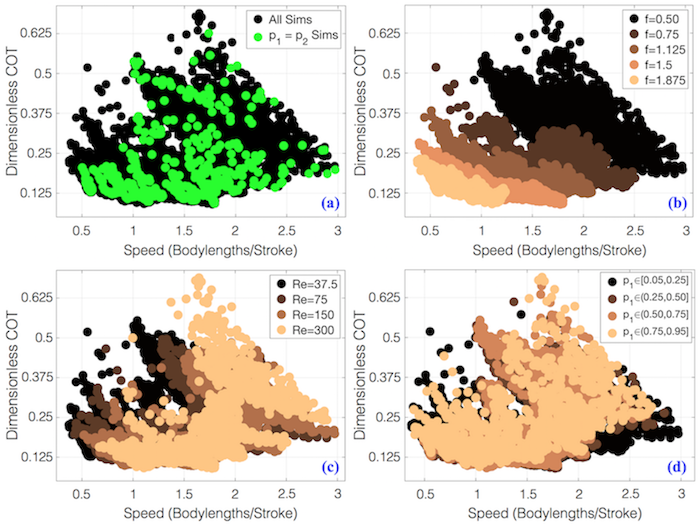}
    \caption{The performance landscape composed of dimensionless COT and speed given for each simulation performed (each $\mathrm{Re}$, $f$, and duty cycle combination). The data was divided by (a) simulations in which $p_1=p_2$, (b) $f$, (c) $\mathrm{Re}$, and (d) $p_1$.}
    \label{fig:Pareto_1}
\end{figure}

\begin{figure}[H]
    \centering
    \includegraphics[width=0.825\textwidth]{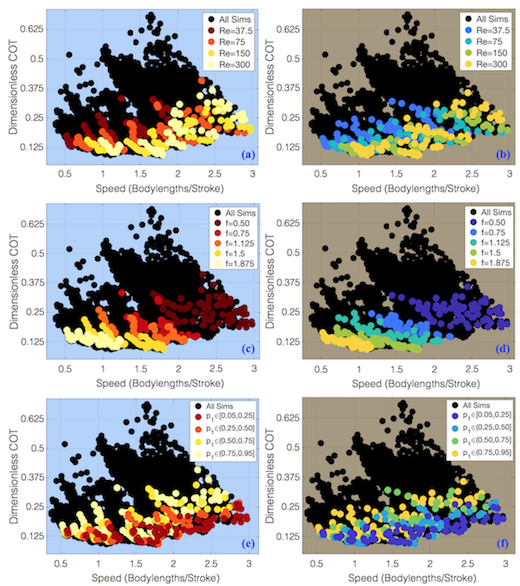}
    \caption{The performance landscape composed of dimensionless COT and speed given for each simulation performed (each $\mathrm{Re}$, $f$, and duty cycle combination). The data was further divided to depict where the optimal combinations from Figures \ref{fig:Percent_Change_Enhance} reside in the performance space. The left and right columns highlights the data partitioned by maximal swimming speed and minimal cost of transport combinations for each $p_1$ and its associated optimal $p_2$, for (a) $\mathrm{Re}$, (b) $f$, and (c) $p_1$.}
    \label{fig:Pareto_2}
\end{figure}

Ideally, we would like to produce optimal duty cycle combinations such that swimming speed is maximized while COT is minimized. Rather than continuing to investigate each locomotion metric independently, we can view the overall performance landscape directly by plotting each jellyfish's speed and COT against each other, see Figures \ref{fig:Pareto_1} and \ref{fig:Pareto_2}. First, Figure \ref{fig:Pareto_1}a gives the overall performance space and highlights the duty cycle combinations in which $p_1=p_2$, that is, no variation in duty cycle from stroke to stroke. These cases are spread throughout the landscape and some even get close to minimizing COT for a given speed. However, since other cases exist that do just that - minimize COT for a given speed, it indicates that varying duty cycles can be beneficial. On the other hand, a poor choice of a different duty cycle combinations can lead to worse performance than swimming with a uniform contraction cycle. Figures \ref{fig:Pareto_1}b, c, and d provide context of where different initial swimming parameters are in the performance space. Figure \ref{fig:Pareto_1}b shows that distinct clusters form in the performance space due to contraction frequency variations, suggesting the performance metrics are highly sensitive to frequency \cite{Battista:ICB2020b,Miles:2019}. Figure \ref{fig:Pareto_1}c suggests that higher Re cases generally result in slightly faster swimming speeds. Figure \ref{fig:Pareto_1}d shows that duty cycles involving lower $p_1$ values could produce either swimmers that fall on the Pareto-optimal curve of minimal COT for maximal speed, or maximal COT for minimal speeds. Figures \ref{fig:Pareto_Layered_Freq}, \ref{fig:Pareto_Layered_Re}, and \ref{fig:Pareto_Layered_p1} in the Supplemental Materials show the explicit regions encompassing particular input values for frequency, Re, or $p_1$, respectively.

To complement Figure \ref{fig:Percent_Change_Enhance}, Figure \ref{fig:Pareto_2} highlights the maximal swimming speed (left) and minimal cost of transport (right) duty cycle combinations for each $\mathrm{Re}$, $f$, and $p_1$ in the performance landscape. Moreover, (a) and (b) group data by $\mathrm{Re}$, (c) and (d) group by $f$, and (e) and (f) by $p_1$. Overall the data further indicates that maximal swimming speeds and maximal cost of transport are not synonymous, nor are minimal swimming speeds and minimal COT. Whether maximizing swimming speed or minimizing COT, both illustrate duty cycle combinations that fall along the Pareto optimal curve, i.e., the curve encompassing the minimal COT for a given swimming speed. Many of the points that fall along the Pareto optimal curve appear to have a duty cycle combination with lower $p_1$ values.

%

%
%

%
%

\section{Discussion and Conclusion}
\label{discussion}

Swimming performance can be tuned by altering different kinematic patterns for variety of locomotive modes. For example, crustaceans use metachronal paddling modes with distinct phase lags to boast more effective, efficient swimming or hovering at certain $\mathrm{Re}$ \cite{Zhang:2014,Ford:2019}. Similarly, fishes use distinct conserved kinematics to better optimize acceleration than when they perform steady state swimming \cite{Akanyeti:2017}. Both of these modes depend on modulating vortex ring formation and interactions, i.e., fluid dynamic processes. In that latter, there is a kinematic transition from initial acceleration to steady state swimming.

Such transitions are commonly observed from aquatic animals for different purposes. For example, when evading predators, larval fish use a fast-start mechanism to increase their likelihood of evading predation \cite{Domenici:1997,Walker:2005}. Or, foraging animals may use different kinematic modes than when steadily swimming \cite{Colin:2003,Carey:2017}. Moreover, a single organism may have different modes for different foraging strategies \cite{Crowder:1985}. In jellyfish, different foraging kinematics and swimming modes are mitigated by morphology \cite{Colin:2002,Colin:2003,Dabiri:2010c}. Although animals display different behavioral kinematics, variations in kinematics from cycle to cycle or their transitional kinematics are not typically investigated.

In this work we performed \textit{in silico} experiments that vary contraction kinematics by combining different duty cycles for a jellyfish that swims via jet propulsion. Two dimensional fluid-structure interaction simulations showed that combining different duty cycles together could lead to enhanced (or impeded) swimming performance. Differences were more pronounced in speed and COT for lower actuation frequencies when comparing cases of two distinct duty cycles combinations to the case of one constant duty cycle (Figures \ref{fig:Colormaps_Speed} and \ref{fig:Colormaps_COT}). Therefore most of the jellyfish detailed in Table \ref{table:duty_cycles} would benefit more from one uniform duty cycle, except maybe the larger \textit{Sarsia tubulosa} that swam at a frequency of $f\sim0.91\ \mathrm{Hz}$. Possible overall enhancements from non-uniform duty cycle kinematics were quantified against cases using one uniform duty cycle (Figure \ref{fig:Percent_Change_Enhance} and Supplemental Figures \ref{fig:Colormaps_Change_Speed}, \ref{fig:Colormaps_Change_COT}). The lowest frequency case ($f=0.5\ \mathrm{Hz})$ corresponded to the highest possible boosts in speed (maximal differences were on the order of $60-80\%$) among all Re cases investigated. Moreover, a similar trend was observed for COT, but with maximal reductions all around $40-50\%$ for all $Re$ and $f=0.5\ \mathrm{Hz}$. Generally lower $f$ offered greater opportunity for increased performance (both speed and cost of transport) from different duty cycle combinations. Furthermore, robust regions of $(p_1,p_2)$ combinations emerged that increased speed and also reduced the cost of transport for lower $f$ and lower $\mathrm{Re}$.

Unfortunately, combining duty cycles could also yield negative consequences (Figure \ref{fig:Percent_Change_Impede}). Poor combinations could produce reductions in speed of over $40\%$ for specific cases of duty cycle combinations for every $\mathrm{Re}$ and almost every $f$ considered. Increases in COT of upwards of over $40\%$ were possible for every $\mathrm{Re}$ and $f$, depending on $p_1$. Moreover, increases in COT over $100\%$ were possible for cases lower Re and lower $p_1$. Thus care must be taken when combining different duty cycle kinematics.

Performance landscapes showed that lower frequency cases could produce Pareto-optimal swimmers (highest speed for lowest COT) or some of the worst swimmers (lowest speed for highest COT), see Figure \ref{fig:Pareto_1}b. Similarly, duty cycle combinations involving a lower $p_1$ valued duty cycle could yield similar behaviors (Figure \ref{fig:Pareto_1}d). Generally higher $Re$ corresponded to faster swimmers across all duty cycle combinations considered (Figure \ref{fig:Pareto_1}b). However, optimal combinations seem to favor at least one duty cycle with lower $p_1$ (Figures \ref{fig:Pareto_2}e and f), as they are closer to the Pareto-optimal curve.

Recently, a study has suggested that jellyfish are able to take advantage of the boost in performance that swimming near a solid boundary provides - the so-called \textit{ground effect} \cite{Gemmell:2021}. Swimming near a solid boundary has shown to boost cruising speeds by upwards of $25\%$ \cite{Fernandez:2015} and thrust by $40-45\%$ \cite{Quinn:2014,Fernandez:2015}. While jellyfish swim in the open ocean and not near solid boundaries, they are able to take advantage an analogous effect by means of the vortex-vortex interactions they produce \cite{Gemmell:2021}. Previously, optimal kinematic parameters, like frequency and duty cycle, have been identified that lead to enhanced performance \cite{Hoover:2015,Miles:2019}. Since different contraction kinematics inherently lead to different vortex-vortex interactions \cite{Alben:2013,Park:2015,Miles:2019,Pallasdies:2019}, this work has illustrated that it is possible to enhance performance via combining optimal duty cycle combinations that lead to beneficial vortex-vortex dynamics between cycles. Entering a new cycle using the same bell kinematics but different starting vortex dynamics leads to different performance. Not only that but the vortex dynamics inherently affect the bell's pressure dynamics, which largely impacts forward propagation \cite{Gemmell:2015}. Therefore part of the enhancement shown in this work is likely due to \textit{virtual wall effects}, as described by Costello et al. 2020 and Gemmell et al. 2021 \cite{Costello:2020,Gemmell:2021}. The virtual wall effect refers to the phenomenon in which fluid at the interface between two oppositely spinning vortices moves faster than if only one of the vortices were present. Similar behavior would be observed if it were only one vortex and a solid wall \cite{Schmid:2009}. While there are unanswered questions to focus on in terms of biological propulsion, i.e., how the vortex-vortex interactions (e.g., virtual wall effects) directly impact swimming efficiency and economy, this study provides a basis for further exploration by suggesting cases in which boosts in performance emerge due to combining different duty cycles, along with their accompanying performance data.

This study did not assess differences in material properties or morphology \cite{Colin:2002,Dabiri:2007,Dabiri:2010b,Colin:2012,Hoover:2015,Katija:2015,Hoover:2019,Miles:2019b}, locomotion mode, i.e., either jetting, jet-paddling/rowing, or transitions there-between \cite{Dabiri:2007,Weston:2009}, or incorporate gliding time between successive cycles \cite{Peng:2012,Akoz:2018}. It also did not investigate the possibility of how combining different contraction kinematics via different frequency combinations, rather than duty cycles, could lead to possible boosts or decays in performance. Future work could explore such effects in tandem with duty cycle combinations. Such information can aid in designing biomimetic jellyfish \cite{Frame:2018,Christianso:2019,Almubarak:2020}. In fact, some robotic jellyfish studies are beginning to explore how kinematics affect performance \cite{Ren:2019}. Further investigation into producing optimal vortex-vortex interactions may lead to new ideas of how to design efficient bioinspired underwater robots that are able to take advantage of yet another fluid dynamics phenomenon to heighten performance.

This study did not assess differences in morphology \cite{Dabiri:2010b,Miles:2019b}, locomotion mode, i.e., either jetting, jet-paddling/rowing, or transitions there-between \cite{Dabiri:2007,Weston:2009}, or incorporate gliding time between successive cycles \cite{Peng:2012,Akoz:2018}. It also did not investigate the possibility of how combining different contraction kinematics via different frequency combinations, rather than duty cycles, could lead to possible boosts or decays in performance. Future work could explore such effects in tandem with duty cycle combinations. Such information can aid in designing biomimetic jellyfish \cite{Frame:2018,Christianso:2019,Almubarak:2020}. In fact, some robotic jellyfish studies are beginning to explore how kinematics affect performance \cite{Ren:2019}. Further investigation into producing optimal vortex-vortex interactions may lead to new ideas of how to design efficient bioinspired underwater robots that are able to take advantage of yet another fluid dynamics phenomenon to heighten performance.

%
%

\section*{Acknowledgements}


The authors would also like to thank Christina Battista, Karen Clark, Jana Gevertz, Laura Miller, Matthew Mizuhara, and Emily Slesinger for comments and discussion. The authors also wish to thank the anonymous reviewers for their careful reading of the manuscript and their very insightful, constructive feedback that strengthened the manuscript. Computational resources were provided by the NSF OAC \#1826915 and the NSF OAC \#1828163. Support for N.A.B. was provided by the TCNJ Support of Scholarly Activity Grant, the TCNJ Department of Mathematics and Statistics, and the TCNJ School of Science.

%
%

\bibliographystyle{elsarticle-num}
\bibliography{Swim}

%
%
%
%
%

\newpage
\setcounter{page}{2}
\setcounter{section}{0}


\beginsupplement

\begin{center}
    \large{\textbf{SUPPLEMENTAL MATERIALS}} \\
\end{center}

\vspace{0.25in}

%
%

\section{Details on IB}
\label{IB_Appendix}

A two-dimensional formulation of the immersed boundary (IB) method is discussed below. The open-source IB software, \textit{IB2d}, was used to perform all fluid-structure interaction simulations \cite{Battista:2015,BattistaIB2d:2017,BattistaIB2d:2018}. The software itself has been previously validated \cite{BattistaIB2d:2017} and specific grid size and resolution convergence tests for this jellyfish model were also previously performed \cite{BattistaMizuhara:2019,Miles:2019}. For a full review of the immersed boundary method, please see Peskin 2002 \cite{Peskin:2002},  Mittal and Iaccarino 2005 \cite{Mittal:2005}, and/or Griffith and Patankar 2020 \cite{Griffith:2020}. 

%
%

\subsection{Governing Equations of IB}

The conservation of momentum and mass equations that govern an incompressible and viscous fluid are listed below:

\begin{equation} 
   \rho\Big[\frac{\partial\u}{\partial t}({\bf x},t) +\u({\bf x},t)\cdot\nabla \u({\bf x},t)\Big]=  -\nabla p({\bf x},t) + \mu \Delta \u({\bf x},t) + \f({\bf x},t) \label{eq:NS1}
 \end{equation}
  \begin{equation}
      \div \u({\bf x},t) = 0 \label{eq:NSDiv1}
  \end{equation}
where $\u({\bf x},t) $ is the fluid velocity, $p({\bf x},t) $ is the pressure, $\f({\bf x},t) $ is the force per unit area applied to the fluid by the immersed boundary, $\rho$ and $\mu$ are the fluid's density and dynamic viscosity, respectively. The independent variables are the time $t$ and the position ${\bf x}$. The variables $\u, p$, and $\f$ are all written in an Eulerian frame on the fixed Cartesian mesh, $\textbf{x}$. 

The interaction equations, which handle all communication between the fluid (Eulerian) grid and immersed boundary (Lagrangian grid) are the following two integral equations:
\begin{align}
   {\bf f}({\bf x},t) &= \int {\bf F}(s,t)  \delta\left({\bf x} - {\bf X}(s,t)\right) ds \label{eq:force1} \\
   {\bf U}({\bf X}(s,t))  &= \int \u({\bf x},t)  \delta\left({\bf x} - {\bf X}(s,t)\right) d{\bf x} \label{eq:force2}
\end{align}
where ${\bf F}(s,t)$ is the force per unit length applied by the boundary to the fluid as a function of Lagrangian position, $s$, and time, $t$, $\delta({\bf x})$ is a two-dimensional delta function, and ${\bf X}(s,t)$ and ${\bf U}(s,t)$ give the Cartesian coordinates and velocity at time $t$ of the material point labeled by the Lagrangian parameter, $s$, respectively. The Lagrangian forcing term, ${\bf F}(s,t)$, gives the deformation forces along the boundary at the Lagrangian parameter, $s$. Equation (\ref{eq:force1}) applies this force from the immersed boundary to the fluid through the external forcing term in Equation (\ref{eq:NS1}). Equation (\ref{eq:force2}) moves the boundary at the local fluid velocity. This enforces the no-slip condition. Each integral transformation uses a two-dimensional Dirac delta function kernel, $\delta$, to convert Lagrangian variables to Eulerian variables and vice versa.

Using delta functions as the kernel in Eqs.(\ref{eq:force1}-\ref{eq:force2}) is what gives IB its power. To approximate these integrals, discretized (and regularized) delta functions are used. We use the ones given from \cite{Peskin:2002}, i.e., $\delta_h(\mathbf{x})$, 
\begin{equation}
\label{delta_h} \delta_h(\mathbf{x}) = \frac{1}{h^3} \phi\left(\frac{x}{h}\right) \phi\left(\frac{y}{h}\right) \phi\left(\frac{z}{h}\right) ,
\end{equation}
where $\phi(r)$ is defined as
\begin{equation}
\label{delta_phi} \phi(r) = \left\{ \begin{array}{l} \frac{1}{8}(3-2|r|+\sqrt{1+4|r|-4r^2} ),\ \ \ \ \ 0\leq |r| < 1 \\    
\frac{1}{8}(5-2|r|+\sqrt{-7+12|r|-4r^2}),\ 1\leq|r|<2 \\
0 \hspace{2.35in} 2\leq |r|.\\
\end{array}\right.
\end{equation}

%
%

\subsection{Numerical Algorithm}
We imposed periodic boundary conditions on the rectangular domain. To solve Equations (\ref{eq:NS1}), (\ref{eq:NSDiv1}),(\ref{eq:force1}) and (\ref{eq:force2}) we needed to update the fluid's velocity and pressure as well as the position of the boundary and force acting on the boundary at time $n+1$ using data from time $n$. The IB does this in the following steps \cite{Peskin:2002}.

\textbf{Step 1:} Find the force density, ${\bf{F}}^{n}$ on the immersed boundary, from the current boundary configuration, ${\bf{X}}^{n}$.\\
\indent\textbf{Step 2:} Use Equation (\ref{eq:force1}) to spread this boundary force from the Lagrangian boundary mesh to the Eulerian fluid lattice points.\\
\indent\textbf{Step 3:} Solve the Navier-Stokes equations, Equations (\ref{eq:NS1}) and (\ref{eq:NSDiv1}), on the Eulerian grid. Upon doing so, we are updating ${\bf{u}}^{n+1}$ and $p^{n+1}$ from ${\bf{u}}^{n}$, $p^{n}$, and ${\bf{f}}^{n}$. \\
\indent\textbf{Step 4:} Update the material positions, ${\bf{X}}^{n+1}$,  using the local fluid velocities, ${\bf{U}}^{n+1}$, computed from ${\bf{u}}^{n+1}$ and Equation (\ref{eq:force2}).
%

%
%

\section{Additional Simulation and Supporting Data}
\label{app:Additional_Data}

This section provides additional supplemental figures to complement the discussions in previous sections of the manuscript.\\

\begin{figure}[H]
    \centering
    \includegraphics[width=0.9\textwidth]{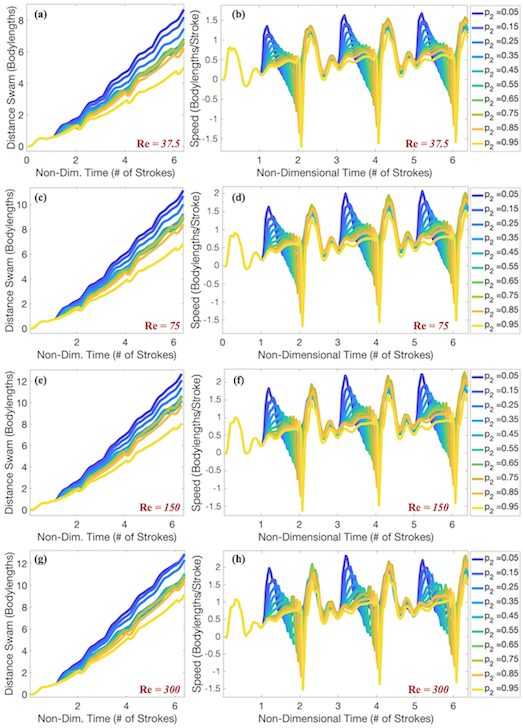}
    \caption{The distances swam (left column) and swimming speeds (right column) over time for cases with $f=0.75\ \mathrm{Hz}$ and $p_1=0.20$, and (a)-(b) $\mathrm{Re}=37.5$, (c)-(d) $\mathrm{Re}=75$, (e)-(f) $\mathrm{Re}=150$, and (g)-(h) $\mathrm{Re}=300$.}
    \label{fig:TimeEvo_f0pt75_VaryRe}
\end{figure}

Figure \ref{fig:TimeEvo_f0pt75_VaryRe} provides temporally data on the distance swam and instantaneous speed of the jellyfish across the first 6 strokes for cases involving $f=0.75\ \mathrm{Hz}$, $p_1=0.20$, and a variety of $\mathrm{Re}$ and $p_2$. Interestingly, as $\mathrm{Re}$ increases, the shape of the speed waveform remains relatively consistent among all cases, unlike that when contraction frequency is varied (see Figure \ref{fig:TimeEvo_Re75_VaryFreq}).

\begin{figure}[H]
    \centering
    \includegraphics[width=0.99\textwidth]{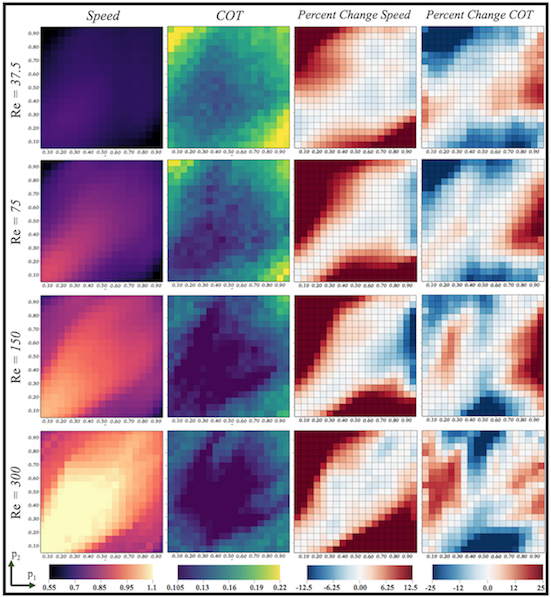}
    \caption{Colormaps representing the time-averaged dimensionless swimming speed and cost of transport as well as the percent change in speed and COT, when compared to the $p1=p_2$ case, for $f=1.875\ \mathrm{Hz}$ and each $\mathrm{Re}$.}
    \label{fig:Colormaps_f1pt875}
\end{figure}

Figure \ref{fig:Colormaps_f1pt875} provides colormaps of performance data (speed and cost of transport) as well as the relative changes in performance data as different duty cycle combinations are considered. As mentioned in Section \ref{results}, at higher frequencies, optimal combinations of duty cycle appear to be those in which $p_1\approx p_2$. However, there do appear to be combinations in which COT is reduced, when compared to the $p_1=p_2$ case. Moreover, branching patterns emerge in the COT colormaps (see $\mathrm{Re}=150$ and $300$ panels) involving duty cycle combinations with minimal COT values.

\begin{figure}[H]
    \centering
    \includegraphics[width=0.55\textwidth]{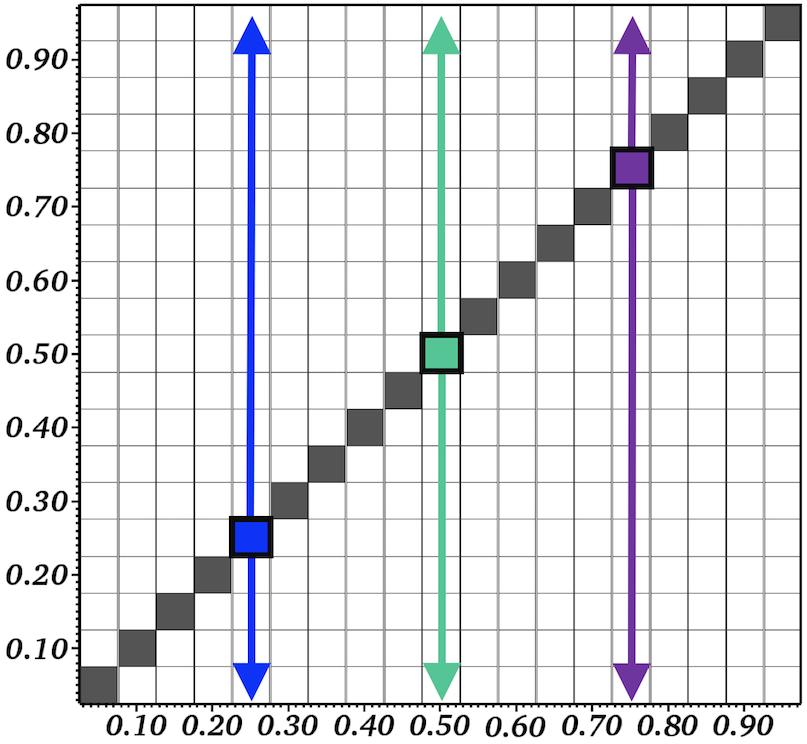}
    \caption{An illustration of which simulations (partitioned by duty cycles) are being compared to compute the percent change in swimming performance.}
    \label{fig:Relative_Change_Idea}
\end{figure}

\begin{figure}[H]
    \centering
    \includegraphics[width=0.99\textwidth]{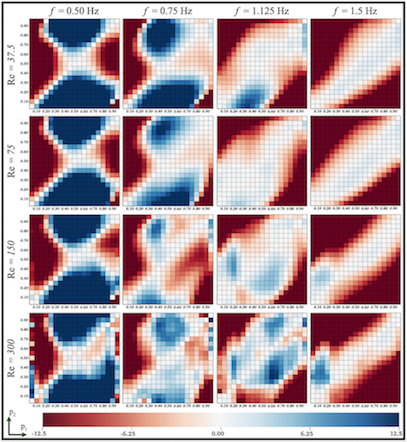}
    \caption{Colormaps representing the percent change in time-averaged dimensionless swimming speed for different duty cycle combinations for each $\mathrm{Re}$ and $f=0.50,0.75,1.125$, and $1.5$ Hz. The case when $f=1.875\ \mathrm{Hz}$ is given in Figure \ref{fig:Colormaps_f1pt875}.}
    \label{fig:Colormaps_Change_Speed}
\end{figure}

\begin{figure}[H]
    \centering
    \includegraphics[width=0.99\textwidth]{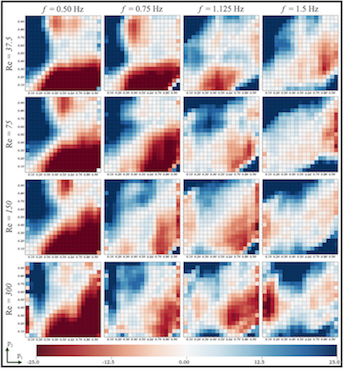}
    \caption{Colormaps representing the percent change in time-averaged cost of transport for different duty cycle combinations for each $\mathrm{Re}$ and $f=0.50,0.75,1.125$, and $1.5$ Hz. The case when $f=1.875\ \mathrm{Hz}$ is given in Figure \ref{fig:Colormaps_f1pt875}.}
    \label{fig:Colormaps_Change_COT}
\end{figure}

Figure \ref{fig:Relative_Change_Idea} shows which simulations were compared when computing the percent changes in performance for Figures \ref{fig:Percent_Change_Enhance}, \ref{fig:Percent_Change_Impede}, \ref{fig:Colormaps_Change_Speed}, \ref{fig:Colormaps_Change_COT}, as well as 
\ref{fig:Colormaps_f1pt875}.

Supplemental Figures \ref{fig:Colormaps_Change_Speed} and \ref{fig:Colormaps_Change_COT} illuminate regions on the $(p_1,p_2)$ space in which led to enhance or impede performance. As suggested previously by Figure \ref{fig:Colormaps_Speed}, the higher frequency cases indicated that combinations in which $p_1\approx p_2$ generally led to faster swimming speeds (right two columns on Figure \ref{fig:Colormaps_Change_Speed}). However, observations were different in the lower frequency cases. That is, for $f=0.50$ and $0.75\ \mathrm{Hz}$, there were robust regions in which $p_1\not\approx p_2$ that led to faster swimming. Moreover, the lower $\mathrm{Re}$ cases also had, both, the most robust regions as well as the greatest positive percent changes. Furthermore, Figure \ref{fig:Colormaps_Change_COT} shows that in these cases ($f=0.5, 0.75\ \mathrm{Hz}$ and $Re=37.5,75$) that the regions of \textit{faster} swimming speed are also associated with regions of \textit{decreased} cost of transport. More generally this observation is consistent across every case considered - higher swimming speed combinations also led to lower a cost of transport as well. From the way in which the cost of transport is calculated (Eq. \ref{eq:COT_dim}), that can be expected to an extent. However, these results indicate that different duty cycle combinations do not require enough additional input power by the bell to negate the reduction in COT through faster swimming. Thus, different duty cycle combinations may benefit both swimming speed and cost of transport (efficiency) in tandem. We took the data from Figures \ref{fig:Colormaps_Change_Speed} and \ref{fig:Colormaps_Change_COT} and computed maximum in each column (maximum positive and negative value for speed and COT, respectively) for the data presented in Figures \ref{fig:Percent_Change_Enhance} and \ref{fig:Percent_Change_Impede}.

\begin{figure}[H]
    \centering
    \includegraphics[width=0.7\textwidth]{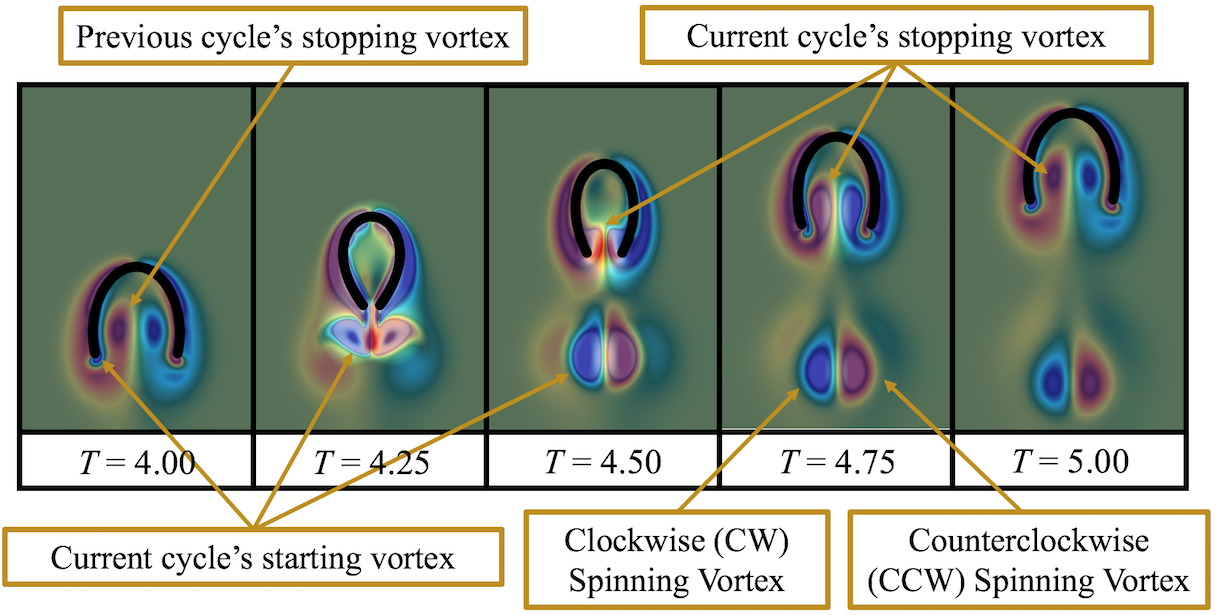}
    \caption{Snapshots of vorticity through the $5^{th}$ contraction cycle, for the case involving $\mathrm{Re}=37.5$, $f=0.50\ \mathrm{Hz}$, and $p_1=p_2=0.2$. The vorticity colormap thresholds in each case are identical across all cases.}
    \label{fig:Vortex_Defn}
\end{figure}

Jet propulsion in jellyfish involves a few key vortices that form and interact throughout a contraction cycle. Figure \ref{fig:Vortex_Defn} outlines the key vortex definitions, formation, and dynamics in a case in which $\mathrm{Re}=75$, $f=0.50\ \mathrm{Hz}$, and $p_1=p_2=0.50$. In short, as the bell contracts, it expels a vortex ring that interacts with the previous cycle's stopping vortex. This interaction is likely where virtual wall effects give rise to boosts in performance. As the jellyfish bell expands, it pulls in the fluid that is entrained from around its bell, in the form of the next stopping vortex. While the stopping vortex may slow its forward movement \cite{Gemmell:2014}, it also provides a mechanism to lower its energetic demand through passive energy recapture \cite{Gemmell:2013}.

\begin{figure}[H]
    \centering
    \includegraphics[width=0.99\textwidth]{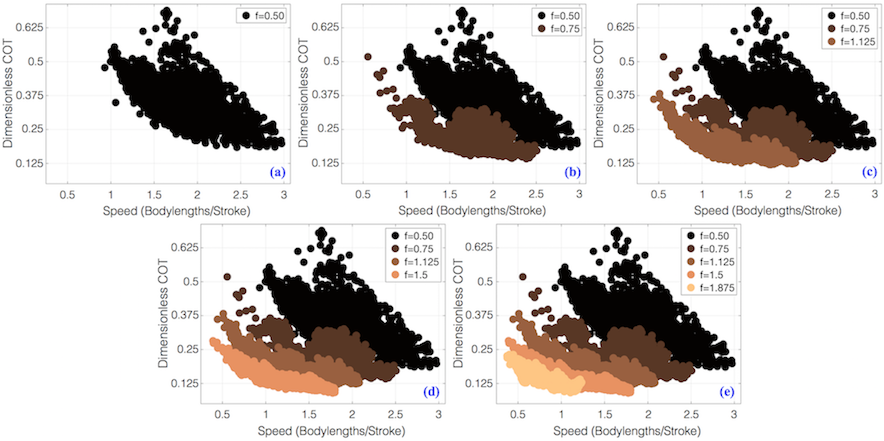}
    \caption{Constructing the overall performance landscape composed of dimensionless COT and speed by layering more of the considered frequencies, $f$.}
    \label{fig:Pareto_Layered_Freq}
\end{figure}

\begin{figure}[H]
    \centering
    \includegraphics[width=0.825\textwidth]{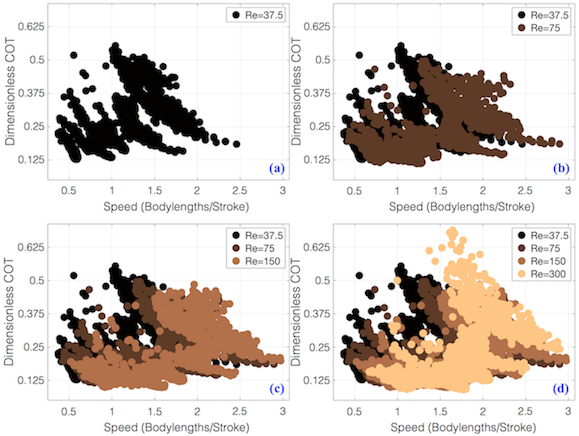}
    \caption{Constructing the overall performance landscape composed of dimensionless COT and speed by layering more of the considered Reynolds numbers, $\mathrm{Re}$.}
    \label{fig:Pareto_Layered_Re}
\end{figure}

\begin{figure}[H]
    \centering
    \includegraphics[width=0.825\textwidth]{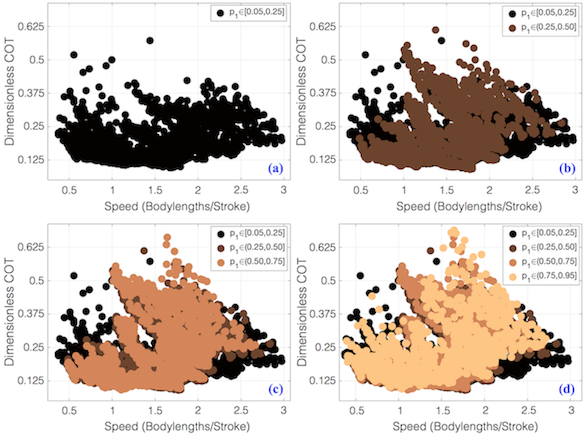}
    \caption{Constructing the overall performance landscape composed of dimensionless COT and speed by layering more ranges of $p_1$ values.}
    \label{fig:Pareto_Layered_p1}
\end{figure}

Figures \ref{fig:Pareto_Layered_Freq}, \ref{fig:Pareto_Layered_Re}, and \ref{fig:Pareto_Layered_p1} provide the deconstructed performance landscapes, as organized by different parameters - contraction frequency ($f$), Reynolds number ($\mathrm{Re}$), and first duty cycle ($p_1$), respectively. As discussed in Section \ref{results}, the performance space is sharply partitioned by frequency. While dimensionless speed is calculated by dividing a simulation's \textit{dimensional} speed by frequency times bell diameter, other locomotion studies have observed this clustering in both \textit{dimensional} and \textit{non-dimensional} data \cite{Battista:ICB2020b,Miles:2019}. Moreover, these studies also suggested that either form of swimming speed is most sensitive to variations in stroke/contraction frequency.  

\end{document}